\documentclass[reprint, amsmath,amssymb,
 aps,
]{revtex4-2}

\usepackage{latexsym}
\usepackage{epsfig}
\usepackage{xcolor}
\usepackage{float}
\usepackage{graphicx}
\usepackage{epstopdf}
\usepackage{dcolumn}
\usepackage{multirow}
\usepackage{colortbl}
\usepackage{subfig}
\usepackage{footnote}
\RequirePackage[colorlinks,citecolor=blue,urlcolor=blue,linkcolor=blue]{hyperref}

\newcolumntype{C}[1]{>{\centering\arraybackslash}m{\dimexpr#1-2\tabcolsep\relax}}
\begin{document}


\title{A Double-Sine-Gordon Early Universe}

\author{Behnoush Afshar$^1$}
 \email{behnoush.afshar.cosmology@gmail.com}
\author{Marziyeh Peyravi$^2$}%
 \email{marziyeh.peyravi@stu-mail.um.ac.ir}

\author{Kazuharu Bamba$^3$}%
\email{bamba@sss.fukushima-u.ac.jp}

\author{Hooman Moradpour$^1$}%
\email{h.moradpour@riaam.ac.ir}
\address{$^1$ Research Institute for Astronomy and Astrophysics of Maragha
(RIAAM), University of Maragheh, P.O. Box 55136-553, Maragheh,
Iran\\
$^2$ Department of Physics, School of Sciences, Ferdowsi University of Mashhad,  91775-1436, Mashhad, Iran\\
$^3$Faculty of Symbiotic Systems Science, Fukushima University, Fukushima 960-1296, Japan}




\begin{abstract}
    A solitonic model of the early universe is introduced by employing the Double-Sine-Gordon (DSG) potential.
    The model predicts the  appropriate number of e-foldings ($N_e$) required for favored inflation and is an advantage for
    the model in addressing the flatness, horizon, and magnetic monopole problems.
    Compatibility of the model with observations, including the Planck $2018$ data \cite{Akrami et al. (2020)} and the Planck $2018$ data+BK$18$+BAO
    \cite{Ade et al. (2021)} paves the way to estimate the model's free parameters. The results generate acceptable and
    proper values for the spectral index ($n_s$) and the tensor-to-scalar ratio ($r$) in agreement with the Planck $2018$ data \cite{Akrami et al. (2020)} and the Planck $2018$ data+BK$18$+BAO \cite{Ade et al. (2021)}. Correspondingly, a consistent description of the reheating era is obtained, yielding positive reheating number of e-foldings ($N_{\mathrm{reh}}$) and reheating final temperature ($T_{\mathrm{reh}}$) from $10^{-2}$ GeV to $10^{16}$ GeV. The inflationary trans-Planckian constraint is also concurrently addressed. Overall, the model seems viable at the inflationary and reheating eras.
\end{abstract}

\maketitle

\section{Introduction}
Although the observation of the cosmic microwave background
radiation (CMB \cite{Penzias and Wilson (1965)}) shed light on the
study of the dark ages of the cosmos \cite{Lazarides (2006)}, the
CMB also seems beneficial to study other parts of the cosmos. As an
example, the lensing of CMB photons leads us to study dark energy
that is responsible for the current accelerating universe first
reported by E. Hubble \cite{Hubble (1929)}. The flatness problem,
as formulated by Dicke and Peebles \cite{Dicke (1970), Dicke and
    Peebles (1979)}, rooted in the fine-tuned density of matter and
energy in the early universe leading to the observed flatness on
large-scale structures \cite{Lazarides (2006), Brawer (1995),
    Carroll (2014), Zucchini (2018), Lozanov (2019), Helbig (2020)},
is one of the basic concerns in understanding the early universe.
The explanation of the homogeneity and the thermal equilibrium of
the CMB \cite{Lazarides (2006), Brawer (1995), Carroll (2014),
    Zucchini (2018)}, related to the horizon problem, forms another
problem \cite{Rindler (1956)}. In addition, the magnetic monopole
problem, first highlighted by Dirac \cite{Dirac (1931)}, concerns
the absence of magnetic monopoles despite theoretical predictions
\cite{Lazarides (2006), Zucchini (2018), Lozanov (2019), Rajantie
    (2012), Zeldovich:1978wj}. To address the mentioned challenges, in $1981$, Guth
\cite{Guth (1981)} introduced the concept of inflation as an
exponentially expansion driven by a scalar field. In $1980$,
Starobinsky \cite{Starobinsky (1980)} had also proposed a
different mechanism for inflation using gravitational effects. It
is worthwhile mentioning that the relationship between the
first-order phase transition of vacuum and the expansion of the
universe has been studied by Sato \cite{Sato (1981)}. Others later
expanded and improved these studies \cite{Linde (1982), Albrecht
    and Steinhardt (1982), Hawking et al. (1982), Linde (1983), Mohammadi:2021wde}.

Generally, inflation with the number of e-foldings ($N_e$) in the
range of $40$ to $70$ \cite{Carroll (2014), Zucchini (2018),Remmen
    and Carroll (2014), Huang (2014), Amin et al. (2015), Amin et al.
    (2016), Osses et al. (2021), Mirtalebian et al. (2021),Zhou et al.
    (2022), Afshar et al. (2022), Afshar et al. (2023)} reduces the
initial curvature of the cosmos, finally solving the flatness
problem \cite{Brawer (1995), Carroll (2014), Zucchini (2018),
    Lozanov (2019), Helbig (2020)}. Therefore, the homogeneity and the
thermal equilibrium of the CMB rooted in the fact that these
regions were in a causal relationship at the beginning of
inflation \cite{Brawer (1995), Carroll (2014), Zucchini (2018),
    Lozanov (2019)}. Moreover, inflation can dilute the density of
magnetic monopoles, making them very difficult to detect
\cite{Zucchini (2018), Lozanov (2019), Rajantie (2012)}. It should
also be noted that the idea of inflation not only overcomes the
problems listed above, but it also has substantial experimental
successes in predicting specific aspects of CMB fluctuations and
provides a mechanism for the origin of large-scale structures
\cite{Weinberg (2008),Odintsov:2023weg}.

The reheating mechanism explains the transfer of energy stored in
the inflaton field into the other degrees of freedom that finally
heats the cosmos. This happens in a phase following the primary
inflationary era and is called the reheating era \cite{Lozanov
(2019), Amin et al. (2015), Mirtalebian et al.
    (2021), Zhou et al. (2022), Afshar et al. (2023), Lidde and Leach
    (2003), Allahverdi et al. (2010), Dai et al. (2014), Cheong et al.
    (2022), Kofman:1994rk, Kofman:1997yn, Bassett:2005xm, Khlebnikov:1996zt, Khlebnikov:1996wr, Allahverdi:2000ss}. Three parameters that describe this period are the
reheating number of e-foldings $N_{\mathrm{reh}}$, the reheating
final temperature $T_{\mathrm{reh}}$, and the reheating effective
equation of state (EoS) $\omega_{\mathrm{reh}}$ \cite{Mirtalebian
et al. (2021), Zhou
    et al. (2022), Afshar et al. (2023), Dai et al. (2014), Cheong et
    al. (2022)}. The amount of expansion that happens between the end
of inflation and the beginning of the radiation-dominated era is
denoted by the reheating number of e-foldings $N_{\mathrm{reh}}$
\cite{Mirtalebian et al. (2021), Zhou et al. (2022), Afshar et al.
    (2023), Dai et al. (2014), Cheong et al. (2022)}. $T_{\mathrm{reh}}$
represents the energy of the particles created due to the
reheating \cite{Amin et al. (2015), Dai et al. (2014), Mirtalebian
    et al. (2021), Allahverdi:2000ss}, and the minimum permissible $T_{\mathrm{reh}}$ is
constrained by the fact that the temperature of Big Bang
nucleosynthesis $T_{\mathrm{BBN}}$ cannot be less than $10^{-2}$
GeV \cite{
    Mirtalebian et al. (2021), Zhou et al. (2022), Afshar et al.
    (2023), Dai et al. (2014), Cheong et al. (2022), Allahverdi:2000ss}. Furthermore, the
maximum value of $T_{\mathrm{reh}}$ ($T_{\mathrm{reh},\max}$)
depends entirely on the inflationary model \cite{Gialamas and
Lahanas (2020), Co et al.
    (2020), Ming (2021), Haro and Saló (2023), Allahverdi:2000ss}.
The decay of inflaton field yields particles linked to the Grand
Unified Theory (GUT) scale (an energy range of approximately
$10^{15}$ GeV to $10^{16}$ GeV). At this energy scale, the weak
nuclear force, the strong nuclear force, the electromagnetic
force, and gravity coalesce into a unified force, and
additionally, the temperatures $T_{\mathrm{reh},\max}$ and
$T_{\mathrm{GUT}}$ exhibit close proximity \cite{Boer
    (1994)}.

The relationship between the pressure and energy density
of the dominant fluid in the reheating phase is described by its
EoS ($\omega_{\mathrm{reh}}$) \cite{Mirtalebian et al. (2021), Zhou et al.
    (2022), Afshar et al. (2023), Dai et al. (2014), Cheong et al.
    (2022)}. In the inflationary era, $\omega<-\frac{1}{3}$ is
acceptable and $\omega=-1$ is considered as most ideal state
\cite{Saha et al. (2020)}. In this regard, it is useful to mention
that the EoS parameters for cosmic strings and domain walls are
$\omega=-\frac{1}{3}$ and $\omega=-\frac{2}{3}$, respectively
\cite{Burgazli et al. (2015)}. Cosmic strings and domain walls
might arise as the topological defects that occurred during a
phase transition in the early universe \cite{Nemiroff and Patla
    (2008)}. When dust becomes the dominant fluid in the Cosmos and thus controls its dynamics, since it is a pressure-less fluid, the matter dominated era is describable by considering $\omega=0$. On the
other hand, radiation (an extremely relativistic source) meets
$\omega=\frac{1}{3}$ \cite{Baumann (2022)}, and thus, particles
with $\omega>\frac{1}{3}$, known as ultra-light particles, have speeds
exceeding that of light and have not been observed thus far
\cite{Nemiroff and Patla (2008)}.

Inflationary models usually contain scalar fields with high energy
density, known as inflaton fields \cite{Mukhanov (2005), Hobson et
    al. (2006), Durrer (2008)}. According to the simplest field
theory, scalar fields interact only with gravitational
interactions, but not with other matter or radiation \cite{Hobson
    et al. (2006), Roos (2003)}. While GUT allows various solitonic
structures, such as cosmic strings, magnetic monopoles and domain
walls \cite{Kibble (1976), Vilenkin and Shellard (1994)},
inflation reduces the number of these solitonic structures
drastically \cite{Vilenkin and Shellard (1994), Riotto and Trodden
    (2002)}. Therefore, it would be constructive to provide and study
solitonic models of inflation. According to these ideas, studies
have been conducted on inflation in the context of Soliton fields
\cite{Freese et al. (1990), Enqvist et al. (2002), Freese and
Kinney (2004), Chen et al. (2005), Engel (2010), Freese and Kinney
(2015), Mielke (2019), Musmarra et al. (2019), Mielke (2020)}. The
papers \cite{Freese et al. (1990), Freese and Kinney (2004),
Freese and Kinney (2015), Mielke (2019), Mielke (2020)} presented
natural inflation models with the pseudo-Nambu-Goldstone bosons
(PNGBs) potential and evaluating their ability to agree
observational data and produce suitable $N_e$. The mass inflation
models \cite{Enqvist et al. (2002)}, brane inflation \cite{Chen et
al. (2005), Engel (2010)}, as well as the relationship between
inflationary models with dark energy and dark matter have also
been studied \cite{Mielke (2019), Musmarra et al. (2019)}.

However, there is a lack of studies on the reheating era and the consistency of the results for $n_s$ and $r$ with the Planck $2018$ data \cite{Akrami et al. (2020)}
and the Planck $2018$ data+BK$18$+BAO \cite{Ade et al. (2021)}. Moreover, solitonic models have achieved remarkable success in the fields of string theory \cite{Duff et
    al. (1995), Hamanaka et al. (2023)}, dark matter \cite{ Mielke (2019), Grobov
    et al. (2015), Bai et al. (2022), Du et al. (2024)},
black holes \cite{Franzin et al. (2018), Clement and Fabbri (2000)}, and even gravitational waves \cite{Verdaguer (2020),
    Croon et al. (2020), Lozanov et al. (2024)}. Thus, a deeper
knowledge of the early universe in the framework of solitonic
models can shed light on these attempts as well as interactions
between early universe and particle physics that finally opens up
new perspectives on the dynamics and the evolution of the universe
together with the formation of cosmic structures. This motivates
us to examine the eras of inflation and reheating in the framework
of the DSG potential.

For a Soliton field $\phi$, the PNGB and DSG potentials are expressed as

\begin{eqnarray}\label{e1}
V(\phi)=m^4\big(1+\cos(\frac{\phi}{f})\big),
\end{eqnarray}

\noindent and 

\begin{equation}\label{e2}
    V(\phi) = -\alpha \cos (N \phi) +\beta \cos (2N \phi),
\end{equation}

\noindent respectively. In Eq.~(\ref{e1}), $m$ and $f$ are two mass scales and this potential
and its corresponding inflationary model has previously been
studied \cite{Freese et al. (1990), Freese and Kinney (2004), Freese and Kinney (2015), Mielke (2019), Mielke (2020), Bassett:2005xm}. $\alpha$, $\beta$, and $N$ are free parameters of DSG potential. For the sake of convenience, the ratio of $\beta$ to $\alpha $ is stored into parameter
$\gamma$ ($\equiv\frac{\beta }{\alpha}$). Various properties of DSG potential have been extensively studied \cite{Campbell et al. (1986), Gani and Kudryavtsev (1998), Dauxios and Peyrard (2006), Quintero et al. (2009)}. Therefore, motivated by previous investigations on the PNGB inflationay models \cite{Freese et al. (1990), Freese and Kinney (2004), Freese and Kinney (2015), Mielke (2019), Mielke (2020)} and inspired by the DSG's form as a generalization to PNGB and the fact that none of the previous works \cite{Campbell et al. (1986), Gani and Kudryavtsev (1998), Dauxios and Peyrard (2006), Quintero et al. (2009)} have examined the inflationary models corresponding to DSG potential, we are going to study the ability of DSG potential in modeling the early universe.

The paper is organized as follows. In Sec.~\ref{sec2}, focusing on the scalar spectral index $n_s$ and
the tensor-to-scalar ratio $r$ and by confronting the theory with
the Planck $2018$ data \cite{Akrami et al. (2020)} and the Planck $2018$ data+BK$18$+BAO \cite{Ade et al. (2021)}, the permissible values of the free
parameters of model are obtained. In this line, the ability of the model in achieving the number of e-foldings $N_e$
from $30$ to $55$ is also investigated. In Sec.~\ref{sec3}, we examine inflationary trans-Planckian constraint and the constraints it imposes on the free parameters. The reheating era is also
studied in the fourth section. A summary is finally provided at the
last section. In this study, it is assumed that $c=\hbar=1$ and
the reduced Planck mass is indicated by $M_{\mathrm{Pl}}=\sqrt{\frac{1}{8
        \pi G}}={2.435\times{10^{18}}}$ GeV.

\section{Inflation}\label{sec2}
\subsection{Calculating  Inflationary Parameters $n_s$, $r$, and $N_e$}

In this section, the slow-roll inflation driven by the DSG
potential (\textcolor{blue}{2}) is discussed. For a homogeneous
and isotropic universe, the Friedmann-Robertson-Walker (FRW)
metric is written as

\begin{equation}\label{e3}
    g_{\mu\nu} dx^\mu dx^\nu\ =-{dt}^2+{a(t)^2}{\delta_{ij}dx}^i{dx}^j,
\end{equation}

\noindent where $a$ is the scale factor and considered $0$ at the
start of Big Bang. The total action is expressed as

\begin{equation}\label{e4}
    S_{tot}={S_{g}}+{S_{\phi}},
\end{equation}

\noindent in which

\begin{equation}\label{e5}
    S_g=M_{\mathrm{Pl}}^{2}\int{{{d}^{4}}x}\sqrt{-g}(\frac{R}{2}),
\end{equation}

\noindent and

\begin{equation}\label{e6}
    S_{\phi}=\int{{{d}^{4}}x}\sqrt{-g}[-\frac{1}{2}{{g}^{\mu\nu
    }}{{\partial}_{\mu}}{{\phi}^{{}}}{{\partial}_{\nu}}\phi -V(\phi)],
\end{equation}

\noindent denote the gravitational action and the Soliton field
term, respectively. Here, $g={g_{\mu\nu}}{g^{\mu\nu}}$,
$R={R_{\mu\nu}}{R^{\mu\nu}}$ represents the Ricci scalar, and
$V(\phi)$ is the potential of the soliton field $\phi$. Using the
total action (\textcolor{blue}{4}), the Friedmann equations are
achieved as

\begin{equation}\label{e7}
    3 M_{\mathrm{Pl}}^{2} H^2=\frac{1}{2} \dot{\phi}^2+V(\phi),
\end{equation}

\noindent and

\begin{equation}\label{e8}
    -M_{\mathrm{Pl}}^{2}(3 H^2+2\dot{H})=\frac{1}{2} \dot{\phi}^2-V(\phi),
\end{equation}

\noindent in which $H=\frac{\dot{a}}{a}$ denotes the Hubble
parameter, and dot indicates the time derivative. In addition, varying the total action (\textcolor{blue}{4}) with respect to $\phi$, one finds

\begin{equation}\label{e9}
    \ddot{\phi }+3H\dot{\phi }+V'(\phi)=0,
\end{equation}

\noindent where $V'(\phi)=\frac{dV(\phi)}{d\phi}$.  The potential slow-roll parameters  $\varepsilon_V$ and $\eta_V$ are defined as \cite{Lazarides (2006), Zucchini (2018), Lozanov (2019), Remmen and Carroll (2014), Amin et al. (2016), Osses et al. (2021), Zhou et al. (2022), Afshar et al. (2022), Afshar et al. (2023), Cheong et al. (2022), Odintsov:2023weg}

\begin{eqnarray}\label{e10}
     && \varepsilon_V= -\frac{\dot{H}}{H^2}= \frac{M_{\mathrm{Pl}}^{2}}{2} \left(\frac{V'(\phi)}{V(\phi)}\right)^2=\nonumber\\&&\frac{{M_{\mathrm{Pl}}^{2}} N^2}{2} \left(\frac{\sin(N \phi) (1-4 \gamma \cos(N \phi))}{-\cos (N \phi)+\gamma \cos (2 N \phi)}\right)^2 ,
\end{eqnarray}

\noindent and

\begin{eqnarray}\label{e11}
    &&\eta_V= \frac {\ddot{\phi}}{H \dot{\phi}}-\frac{\dot{H}}{H^2}=M_{\mathrm{Pl}}^{2} \frac{V''(\phi)}{V(\phi)}= \nonumber\\&& M_{\mathrm{Pl}}^2 N^2 \frac{\cos (N \phi)-4 \gamma \cos (2N \phi)}{-\cos (N \phi)+\gamma \cos (2N \phi)},
\end{eqnarray}

\noindent and slow-roll inflation requires ${\varepsilon}_V$, $|{\eta}_V|\ll 1$ \cite{Zucchini (2018),
    Lozanov (2019), Remmen and Carroll (2014), Amin et al. (2016),
    Rasanen and Tomberg (2019), Osses et al. (2021), Cheong et al.
    (2022), Afshar et al. (2022), Afshar et al. (2023), Odintsov:2023weg}. The number of
e-foldings is also calculated as \cite{Remmen and Carroll (2014),
    Huang (2014), Amin et al. (2016), Osses et al. (2021), Afshar et
    al. (2022), Afshar et al. (2023)}

\begin{equation}\label{e12}
    N_e= \int_{a_\mathrm{k}}^{a_\mathrm{f}}d\ln a=\int_{t_\mathrm{k}}^{t_\mathrm{f}} {H dt}.
\end{equation}

\noindent Using the relation

\begin{equation}\label{e13}
    H dt=\frac{H}{\dot{\phi}} d\phi=  \frac{|d\phi|}{M_{\mathrm{Pl}}}\frac{1}{\sqrt{2{\varepsilon_V}}},
\end{equation}

\noindent along with  $H^2=-\frac{\dot{H}}{\varepsilon_V}=\frac{\dot{\phi}^2}{2 M_\mathrm{Pl}^2 \varepsilon_V}$, obtained by employing Eqs. (\ref{e7}) and (\ref{e8}), one reaches at

\begin{eqnarray}\label{e14}
 && N_e= \frac{1}{M_{\mathrm{Pl}}^2}\bigg|\int^{\phi_\mathrm{f}}_{\phi_\mathrm{k}} d\phi \frac{V(\phi)}{|V'(\phi)|}\bigg|=  \\&&
\bigg| \frac{1}{2 M_{\mathrm{Pl}}^2 N^2 (-1+16 \gamma^2)} \bigg[sign(-V'(\phi_\mathrm{f}))\big[(1+8 \gamma^2) \nonumber\\&&\ln(|1-4 \gamma \cos ( N \phi_\mathrm{f})|)+2 (-1-3\gamma+ 4\gamma^2) \ln (\sin(\frac{N \phi_\mathrm{f}}{2}))\nonumber\\&&+2(-1+ 3 \gamma+ 4 \gamma^2) \ln (\cos (\frac{N \phi_\mathrm{f}}{2})) \big]- sign(-V'(\phi_\mathrm{k}))\nonumber\\&&\big[(1+8 \gamma^2)\ln(|1-4 \gamma \cos ( N \phi_\mathrm{k})|)+2 (-1-3\gamma+ 4\gamma^2)\nonumber\\&&\ln (\sin(\frac{N \phi_\mathrm{k}}{2}))+2(-1+ 3 \gamma+ 4 \gamma^2) \ln (\cos (\frac{N \phi_\mathrm{k}}{2}))\big]\bigg] \bigg|.\nonumber
\end{eqnarray}

\begin{figure*}
\centering  

    \includegraphics[width=0.45\textwidth]{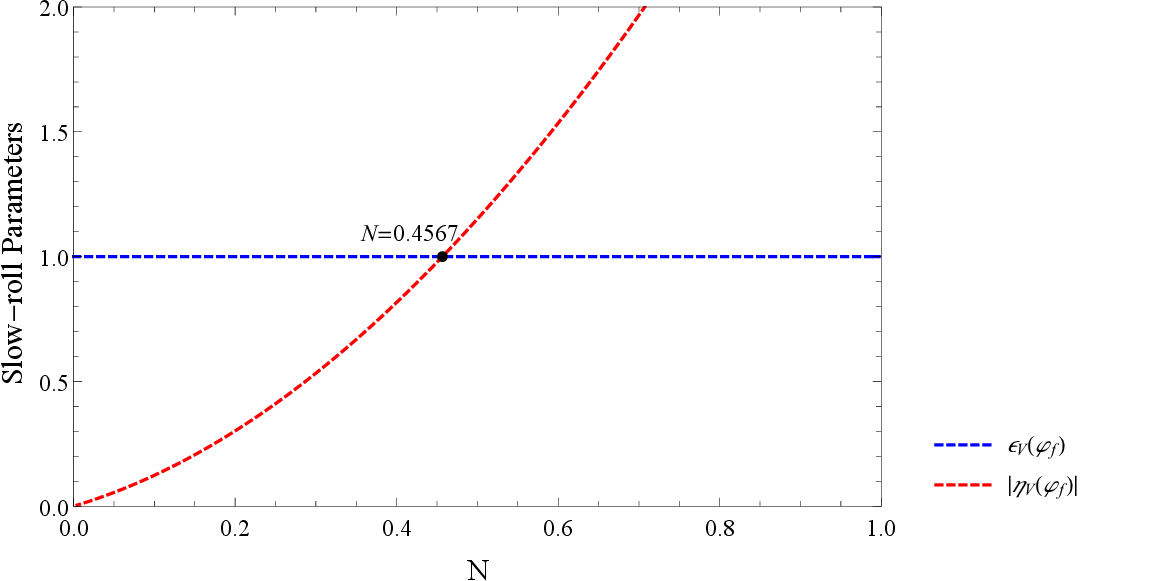} \hspace{-0.3cm}
    \includegraphics[width=0.45\textwidth]{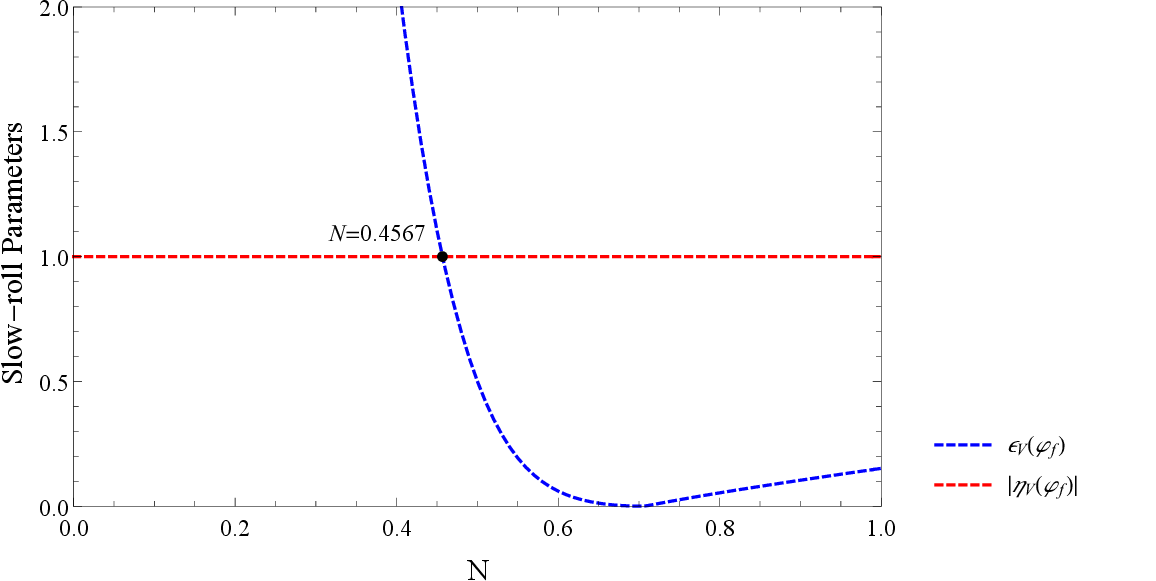}
    \caption{Potential slow-roll parameters ${\varepsilon _V} ({\phi _\mathrm{f}})$ and $|{\eta_V} ({\phi _\mathrm{f}})|$ as functions of $N$ for $\alpha>0$ and $\gamma =-0.5$. The condition for end of inflation in the interval $N\le 0.4567$ ($N\ge 0.4567$) is ${\varepsilon_V} ({\phi _\mathrm{f}}) =1$ ($|{\eta _V} ({\phi _\mathrm{f}})|=1$).}\label{fig1}
\end{figure*}

\begin{figure*}[htp!]

    \centering
     \includegraphics[width=0.45\textwidth]{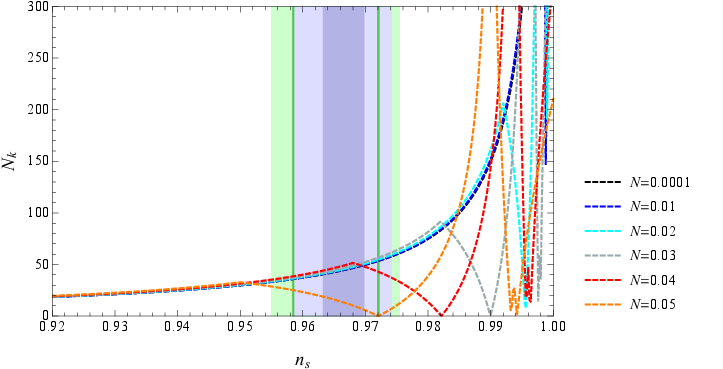}
    \includegraphics[width=0.45\textwidth]{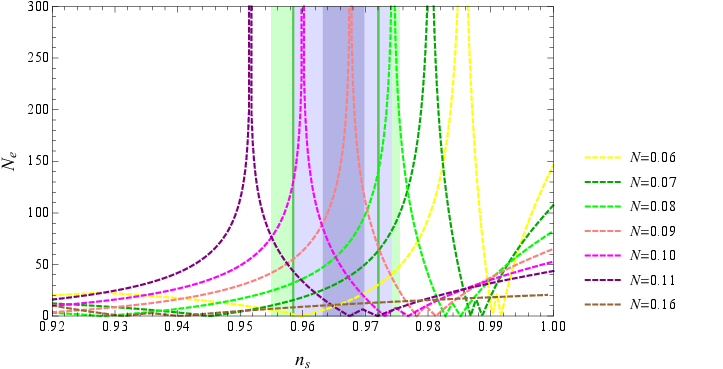}\\
    \caption{Number of e-foldings $N_e$ as a function of the spectral index $n_s$ for the DSG
        potential (\textcolor{blue}{2}), $\alpha>0$, $\gamma=-0.5$, and several
        values of $N$. The green and blue regions represent the
        permissible ranges of $n_s$ based on the Planck $2018$ data \cite{Akrami et al. (2020)} and
        the Planck $2018$ data+BK$18$+BAO \cite{Ade et al. (2021)},
        respectively. The $68\%$ regions are highlighted compared to the $95\%$ regions \cite{Akrami et al. (2020), Ade et al. (2021)}. In this paper, $N_e=30$ is the minimum allowed value. Accordingly, the values of $N > 0.1415$ (such as $N = 0.16$) are considered invalid since they cannot produce $N_e \ge 30$. The curve of $N_e(n_s)$ for a given $N$ is divided into seven branches. For $30 \leq N_e \leq 55$, two of them are incompatible with the Planck $2018$ data \cite{Akrami et al. (2020)} and the Planck $2018$ data+BK$18$+BAO \cite{Ade et al. (2021)}, which makes them unacceptable. The remaining five branches are linked to cases $1–5$ outlined in Table \ref{tab1}. Note that for certain values of $N$, some branches lie within the interval $n_s \le 0.92$. In addition, some branches may become very close to one another near $n_s=1$.}\label{fig2}
\end{figure*}

\noindent Here, the subscripts ``$\mathrm{k}$" and ``$\mathrm{f}$" refer to the Hubble
horizon crossing and the end of inflation, respectively
\cite{Osses et al. (2021)}.  Eq. (\ref{e14}) can be solved for a fixed $N_e$ provided that it is decomposed using the signs of $1- 4\gamma \cos(N\phi)$ and $V'(\phi)$. Furthermore, $\gamma=\pm 0.25$ leads to an infinite $N_e$, and are hence inappropriate. Inflation ends when $\phi_\mathrm{f}$ meets the condition
\cite{Amin et al. (2016)}

\begin{equation}\label{e15}
    \max \big({{\varepsilon_V} ({\phi_\mathrm{f}}),|\eta_{V}({\phi_\mathrm{f}})|}\big)
    =1.
\end{equation}

\begin{figure*}[htp!]

    \centering
    \includegraphics[width=0.45\textwidth]{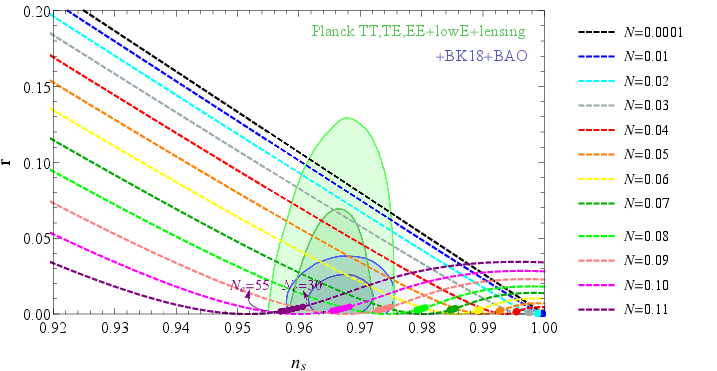} 
    \quad
    \includegraphics[width=0.45\textwidth]{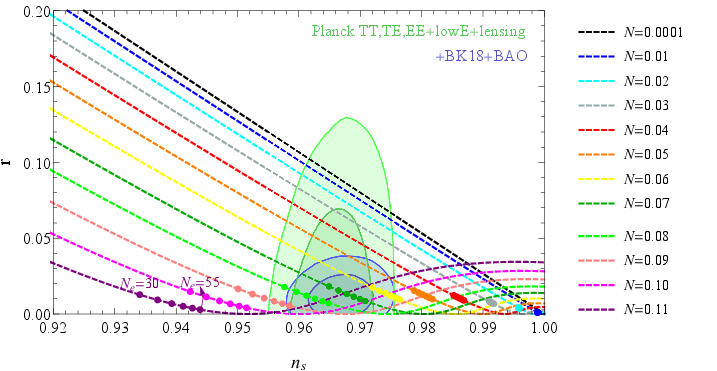} 
    \includegraphics[width=0.45\textwidth]{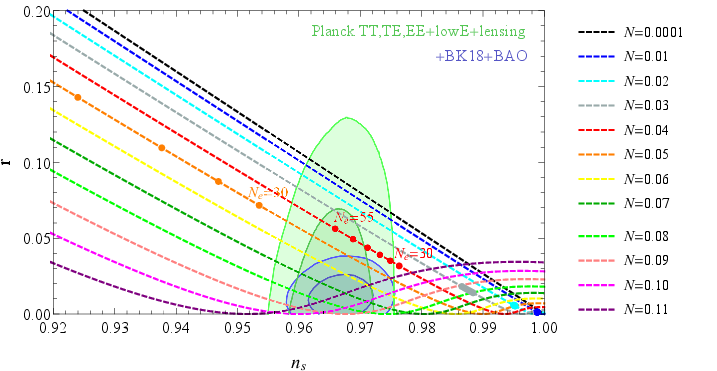}
    \quad
    \includegraphics[width=0.45\textwidth]{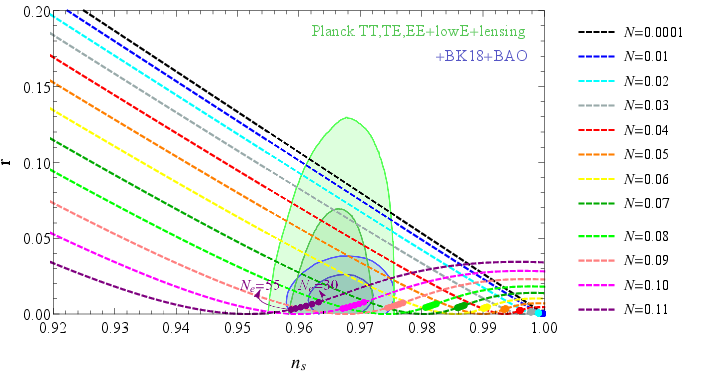}
    \includegraphics[width=0.45\textwidth]{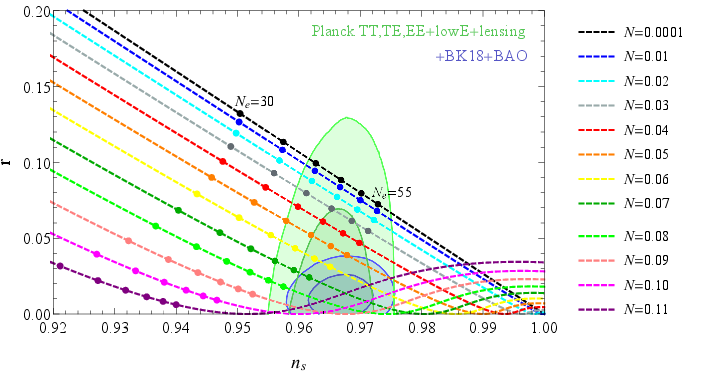} \vspace{0.4cm}
    \caption{
    Tensor-to-scalar ratio $r$ as a function of the spectral index $n_s$ for the DSG potential (\textcolor{blue}{2}), cases $1-5$ mentioned in Table \ref{tab1}, $\alpha>0$, $\gamma=-0.5$, and several values of $N$. The $68\%$ CL and $95\%$ CL of the Planck $2018$ TT,TE,EE+lowE+lensing (green contours) and the Planck $2018$ TT,TE,EE+lowE+lensing+BK$18$+BAO (blue contours) are illustrated, such that the darker inner and lighter outer contours correspond to the $68\%$ CL and $95\%$ CL, respectively \cite{Akrami et al. (2020), Ade et al. (2021)}. The top left (right) sub-figure represents case $1$ (case $2$), the second row, the left (right) sub-figure represents case $3$ (case $4$), and the final sub-figure represents the fifth case. Each $r(n_s)$ curve shows the points $(n_s,r)$ for the constant $N_e$ ranging from $30$ to $55$  with a step size of $5$. Note that in  some cases, for certain values of $N$, these points are very close together. Furthermore, some of the points $(n_s,r)$ in case $3$ have $n_s \le 0.92$, so they are not shown. For each fixed value of $N$, the maximum point on the $N_e(n_s)$ curve aligns with the minimum point on the $r(n_s)$ curve, with the coordinate $(n_{s,m}, r_m)$. The value $n_{s,m} \approx 1$ corresponds to $N = 0.005$, while $n_{s,m} = 0.92$ corresponds to $N = 0.1411$. Therefore, within the interval $0.92 \leq n_s \leq 1$, $r(n_s)$ decreases for $N < 0.005$ and increases for $N > 0.1411$. When $0.005<N<0.1411$, $r(n_s)$ exhibits a minimum extremum at $n_{s,m}$, such that it decreases over $0.92 \leq n_s < n_{s,m}$, and increases over $n_{s,m} < n_s \leq 1$. The Planck $2018$ TT,TE,EE+lowE+lensing \cite{Akrami et al. (2020)} also indicate that the maximum permissible value of $r$ is $0.1295$ \cite{Akrami et al. (2020)}, hence $N > 0.217$  is not permitted. In cases $1$ and $4$, both $r$ and $n_s$ decrease as $N_e$ increases for any fixed value of $N$. In cases $2$ and $5$, $r$ decreases while $n_s$ increases. In contrast, in case $3$, $r$ increases and $n_s$ decreases. Moreover, increasing $N$ for a selected $N_e$ always results in a decrease in $n_s$, while $r$ increases (decreases) in case $3$ (case $5$). In the other cases, the behavior of $r$ with respect to $N$ changes at a specific $N$, denoted as $N*$, where $r$ increases when $N < N*$ and decreases when $N > N*$. The values of $N*$ for cases $1$, $2$, and $4$ and $N_e$ ranging from $30$ to $55$ with a step size of $5$ are provided in Table \ref{tab2}.  Moreover, as shown in Table \ref{tab4}, the intersection of $r(n_s)$ curves for constant values of $N_e$  and $\gamma$ with the $95\%$ CL of the Planck $2018$ data \cite{Akrami et al. (2020)} provide intervals of $N$ that are consistent with the data. As a final note, the availability of valid solutions to Eq.~(\ref{e14}) depends on the choice of free parameters. For example, in case $3$ with $\gamma = -0.5$, no solutions are found when $N > 0.074$ for $N_e = 30$, and when $N > 0.0547$ for $N_e = 55$. Additionally, the value $N = 10^{-4}$ is valid only in case $5$.}\label{fig3}
\end{figure*}

\noindent It is noteworthy that if $\varepsilon_V$ ($|\eta _{V}|$)
touches one before $|\eta _{V}|$ ($\varepsilon_V$), then we have ${\varepsilon_V}
({\phi_\mathrm{f}})=1$ (${|\eta_{V}({\phi_\mathrm{f}})|=1)}$ as the condition for
reaching at the end of inflation. It should be noted that
the two conditions can be satisfied simultaneously for some modified theories of gravity like Refs. \cite{Afshar et al. (2022), Afshar et al. (2023)} and also $V(\phi)\sim\phi ^{2}$  in General Relativity  \cite{Baumann (2022)}. Finally, the scalar spectral index $n_s$ and the tensor-to-scalar ratio $r$ are defined as \cite{Zucchini (2018), Huang (2014), Amin et al. (2016), Osses et al. (2021), Zhou et al. (2022), Cheong et al. (2022), Afshar et al. (2022), Afshar et al. (2023), Odintsov:2023weg}

\begin{eqnarray}\label{e16}
   && {n_s}(\phi_\mathrm{k})=1-6\ {\varepsilon_V}(\phi_\mathrm{k})+2\ {\eta_V}(\phi_\mathrm{k})=
    \nonumber\\&&1-3 M_{\mathrm{Pl}}^2 N^2 \left(\frac{ \sin(N \phi_\mathrm{k}) (1-4 \gamma \cos(N \phi_\mathrm{k}))}{-\cos (N \phi_\mathrm{k})+\gamma \cos (2 N \phi_\mathrm{k})}\right)^2
    \\ &&+2 M_{\mathrm{Pl}}^2 N^2\frac{\cos (N \phi_\mathrm{k}) -4 \gamma \cos (2 N \phi_\mathrm{k})}{-\cos (N \phi_\mathrm{k})+\gamma \cos (2 N \phi_\mathrm{k})},\nonumber
\end{eqnarray}

\noindent and

\begin{eqnarray}\label{e17}
   &&r=16\ {\varepsilon_V}(\phi_\mathrm{k})=
   \\&&8 {M_{\mathrm{Pl}}^{2}} N^2 \left(\frac{\sin(N \phi_\mathrm{k}) (1-4 \gamma \cos(N \phi_\mathrm{k}))}{-\cos (N \phi_\mathrm{k})+\gamma \cos (2 N \phi_\mathrm{k})}\right)^2\hspace{-0.1cm}, \nonumber
\end{eqnarray}

\begin{table}
\centering \caption{Five cases that produce values of $n_s$ and $r$ in agreement with the Planck $2018$ data \cite{Akrami et al. (2020)} and the Planck $2018$ data+BK$18$+BAO \cite{Ade et al. (2021)} are defined by decomposing Eq. (\ref{e14}) according to the signs of the functions $1 - 4\gamma \cos(N\phi)$ and $V'(\phi)$. Since the condition $V(\phi_\mathrm{f}) > 0$ requires that $1 - 4\gamma \cos(N\phi_\mathrm{f}) > 0$, Eq. (\ref{e14}) gives rise to eight possible forms. Among these, two scenarios in which $V'(\phi_\mathrm{k}) > 0$ and $1 - 4\gamma \cos(N\phi_\mathrm{k}) < 0$ are found to be inconsistent with the Planck $2018$ data \cite{Akrami et al. (2020)} and the Planck $2018$ data+BK$18$+BAO \cite{Ade et al. (2021)}. Additionally, $n_s$ diverges to $-\infty$ while $r$ diverges to $\infty$ in another one of the eight possible cases where both $V'(\phi) > 0$ and $1 - 4\gamma \cos(N\phi) > 0$ hold during inflation. In the third subsection of Sec. \ref{sec2}, it is explained that cases $1-5$ become inadmissible in certain ranges of $\gamma$ due to yielding imaginary solutions to Eq. (\ref{e14}), $V(\phi_\text{k})<0$, or incompatibility with the Planck $2018$ data \cite{Akrami et al. (2020)} and the Planck $2018$ data+BK$18$+BAO \cite{Ade et al. (2021)}.} \label{tab1}
\renewcommand{\arraystretch}{1.5}
\begin{tabular}{|c|c|c|c|}
\hline

Case & $V'(\phi_\mathrm{f})$& $V'(\phi_\mathrm{k})$& $1-4 \gamma\cos(N\phi_\mathrm{k})$\\ \hline
1 & Positive & Negative & Negative\\
2 &  & Negative & Positive\\\hline
3 &  & Positive & Positive\\
4 & Negative & Negative & Negative\\
5 & & Negative & Positive\\ \hline
\end{tabular}
\end{table}

\noindent respectively. Using Eq.~(\ref{e16}), one is able to find
the possible values of $\phi_\mathrm{k}$ in the chosen interval $0.92 \le n_{s} \le 1$.

\subsection{Examining Consistency with the Planck Data for Constant $\gamma$ and Variable $N$}

This subsection examines the compatibility of the DSG model with the Planck $2018$ data \cite{Akrami et al. (2020)} and the Planck $2018$ data+BK$18$+BAO \cite{Ade et al. (2021)} for variable $N$, $\alpha>0$, and $\gamma=-0.5$ (the reason for choosing this particular value shall be explained in the next subsection). Answers suggesting a contracting universe are deemed unacceptable. Furthermore, since a negative potential violates the slow-roll condition ($\dot{\phi}^{2} << V(\phi)$), the potential must be positive \cite{Remmen and Carroll (2014), Amin et al. (2016), Zucchini (2018), Odintsov:2023weg}. Because $\alpha<0$ leads to negative potential or imaginary solutions to Eq.~(\ref{e14}), $\alpha>0$ is the only admissible value. In the following, $N>0$ is considered. However, the same results can be obtained by setting $N<0$ and converting $\phi$ to $-\phi$, as $\cos(-x) = \cos(x)$.

As a first step, the condition for the end of inflation (\ref{e15}) is examined. Fig. \ref{fig1} shows $\varepsilon_{V} (\phi_\mathrm{f})$ and
$|\eta_{V} (\phi_\mathrm{f})|$ for $\gamma =-0.5$. The point $N
=0.4567$ corresponds to a scenario where $\varepsilon _{V} (\phi
_\mathrm{f}) =|\eta _{V} (\phi _\mathrm{f})|=1$. For $N\le 0.4567$
and $N\ge 0.4567$, inflation ends when $\varepsilon_{V}
({\phi_\mathrm{f}})=1$ and $|\eta _{V} (\phi _\mathrm{f})|=1$,
respectively. Furthermore, the sign of the expression $1-4\gamma \cos(N\phi_\mathrm{f})$ must be positive to ensure $V(\phi_\mathrm{f})>0$.

In the following analysis, the model's ability to achieve a suitable $N_e$ for addressing the flatness \cite{Dicke (1970),Dicke and Peebles (1979)}, horizon \cite{Rindler (1956)}, and magnetic monopole problems \cite{Dirac (1931)} is assessed. For this purpose, the function $N_e(n_s)$ for $\gamma=-0.5$ and several values of $N$ is shown in Fig. \ref{fig2}. In this paper, the minimum acceptable value of $N_e$ is taken to be $30$ \cite{Afshar et al. (2022), Afshar et al. (2023)}. Additionally, as will be discussed in Sec. \ref{sec4}, for $N_e \geq 56.74$, $N_\text{reh}$ becomes negative. Therefore, considering $55$ as the maximum of $N_e$ seems a reasonable choice. As can be observed from Fig. \ref{fig2}, $N> 0.1415$ is not suitable, as it fails to produce $N_e \ge 30$.

To solve Eq. (\ref{e14}) for a fixed $N_e$ and $\gamma = -0.5$, the equation must first be decomposed based on the signs of the functions $1 - 4\gamma \cos(N\phi)$ and $V'(\phi)$. Since $V(\phi_\mathrm{f}) > 0$ requires $1 - 4\gamma \cos(N\phi_\mathrm{f}) > 0$, Eq. (\ref{e14}) splits into eight possible cases. The case where $V'(\phi) > 0$ and $1 - 4\gamma \cos(N\phi) > 0$ during inflation leads to $n_s \to -\infty$ and $r \to \infty$, which makes this case is not viable. The remaining seven cases correspond to seven distinct branches in the $N_e(n_s)$ curve for each $N$, as shown in Fig. \ref{fig2}. Among these, two scenarios with $V'(\phi_\mathrm{k}) > 0$ and $1 - 4\gamma \cos(N\phi_\mathrm{k}) < 0$ are incompatible with the Planck $2018$ data \cite{Akrami et al. (2020)} and the Planck $2018$+BK$18$+BAO data \cite{Ade et al. (2021)}. Five other possible scenarios are listed in Table \ref{tab1}, hereafter referred to as cases $1–5$.

Fig. \ref{fig3} displays the $r(n_s)$ curves for $\gamma = -0.5$ and several values of $N$ within the interval $0.92 \leq n_s \leq 1$. For any given $N$, the maximum point on the $N_e(n_s)$ curve corresponds to the same $n_s$ value as the minimum point on the $r(n_s)$ curve, with the coordinate $(n_{s,m}, r_m)$. $n_{s,m} \approx 1$ is associated with $N = 0.005$, and $n_{s,m} = 0.92$ is associated with $N = 0.1411$. Thus, $r(n_s)$ decreases when $N < 0.005$ and increases when $N > 0.1411$ over the interval $0.92 \leq n_s \leq 1$. For $0.005<N<0.1411$, the curve of $r(n_s)$ has a minimum extremum at $n_{s,m}$, where it decreases in the range $0.92\le n_{s} <n_{s,m} $, then increases in the range $n_{s,m}<n_{s} \le 1$. As an example, for $N=0.1$,
$r(n_{s})$ descends when $0.92\le n_{s}<0.96$ and ascends when $0.96< n_{s} \le 1$. Note that for each $r(n_s)$ curve, the sign of
$1 - 4\gamma \cos(N\phi_\mathrm{k})$ is positive for $n_s < n_{s,m}$ and negative for $n_s > n_{s,m}$. Additionally, Fig. \ref{fig3} illustrates that the tensor-to-scalar ratio has an upper limit of $r=0.1295$ due to the $95\%$ confidence level (CL) of the Planck $2018$ data \cite{Akrami et al. (2020)}. By analysing the intersection of this upper limit with the $r(n_s)$ curves, it is found that for $N > 0.217$, none of the curves intersect this upper limit line, indicating that these values are inconsistent with the Planck $2018$ data \cite{Akrami et al. (2020)} and the Planck $2018$ data+BK$18$+BAO \cite{Ade et al. (2021)}.

\begin{table}
\centering \caption{Values of $N*$ for cases $1$, $2$ and $4$ and $N_e$ from $30$ to $55$  with a step size of $5$. As can be seen in Fig. \ref{fig3}, $r$ increases in the interval $N<N*$, while it decreases in the interval $N>N*$.}\label{tab2}
\renewcommand{\arraystretch}{1.5}
\begin{tabular}{|c|c|c|c|}
\hline
$N_e$&\multicolumn{3}{|c|}{$N*$ for}\\
  & Case 1& Case 2  & Case 4\\ \hline
30   & 0.1033 & 0.0711    & 0.1064 \\
35   & 0.0956 & 0.0658   & 0.0986 \\
40   & 0.0894 & 0.0615   & 0.0923 \\
45   & 0.0842 & 0.058    & 0.087  \\
50   & 0.0799 & 0.055    & 0.0826 \\
55   & 0.0762 & 0.0524  & 0.0787 \\\hline
\end{tabular}

\end{table}
\normalsize 

Fig. \ref{fig3} also shows the points ($n_s,r$) for cases $1-5$ and $N_e$ from $30$ to $55$ with a step size of $5$. These points highlight several important aspects of the scenarios presented in Table \ref{tab1}. First, the range of $N$ that is consistent with the particular observational data, represented by the specific contour, is determined by computing the intersection points between the $r(n_s)$ curve (associated with one of cases $1-5$ and fixed values of $N_e$ and $\gamma$) and that contour. Table \ref{tab4} summarizes the ranges of $N$ that are compatible with the $95\%$ CL of the Planck $2018$ data \cite{Akrami et al. (2020)} for cases $1-5$, $\gamma = -0.5$, and $ 30 \le N_e \le 55$ with a step size of $5$ (the table also includes additional information related to the reheating phase, the details of which are discussed in Sec. \ref{sec4}). This table demonstrates that the previously established limitations, $N < 0.1415$ and $N < 0.217$ derived from the $N_e(n_s)$ and $r(n_s)$ curves, are become further constrained for each case. It can also be evident that the allowed ranges of $N$ for the $68\%$ CL of the Planck $2018$ data \cite{Akrami et al. (2020)} and the two contours of the Planck $2018$ data+BK$18$+BAO \cite{Ade et al. (2021)} are generally more restricted than those in Table \ref{tab4}; however, depending on the specific data and contour, overlap with the allowed ranges may or may not occur. Moreover, the existence of valid solutions to Eq. (\ref{e14}) varies with the free parameters. For instance, in case $3$ with $\gamma = -0.5$, no solution is found for $N_e = 30$ when $N > 0.074$, and for $N_e = 55$ when $N > 0.0547$. Furthermore, only case $5$ allows a valid solution for $N=10^{-4}$.

As a final note, by increasing $N_e$, for each selected $N$, $r$ and $n_s$ decrease in cases $1$ and $4$, while $r$ decreases and $n_s$ increases in cases $2$ and $5$. In contrast, $r$ increases and $n_s$ decreases in case $3$. Additionally, by increasing $N$, for each selected $N_e$, $n_s$ always decreases while $r$ increases (decreases) in case $3$
(case $5$). In other cases, the behavior of $r$ with respect to $N$ changes at a specific $N$, which is denoted as $N*$, so that $r$ increases (decreases) in the interval $N<N*$ ($N>N*$). The values of $N*$ for cases $1$, $2$, and $4$ and $N_e$ from $30$ to $55$ with a step size of $5$ are mentioned in Table \ref{tab2}.

\subsection{Examining Consistency with the Planck Data for Constant $N$ and Variable $\gamma$}  

\begin{figure*}[htp!]
    \centering
    \includegraphics[width=0.45\textwidth]{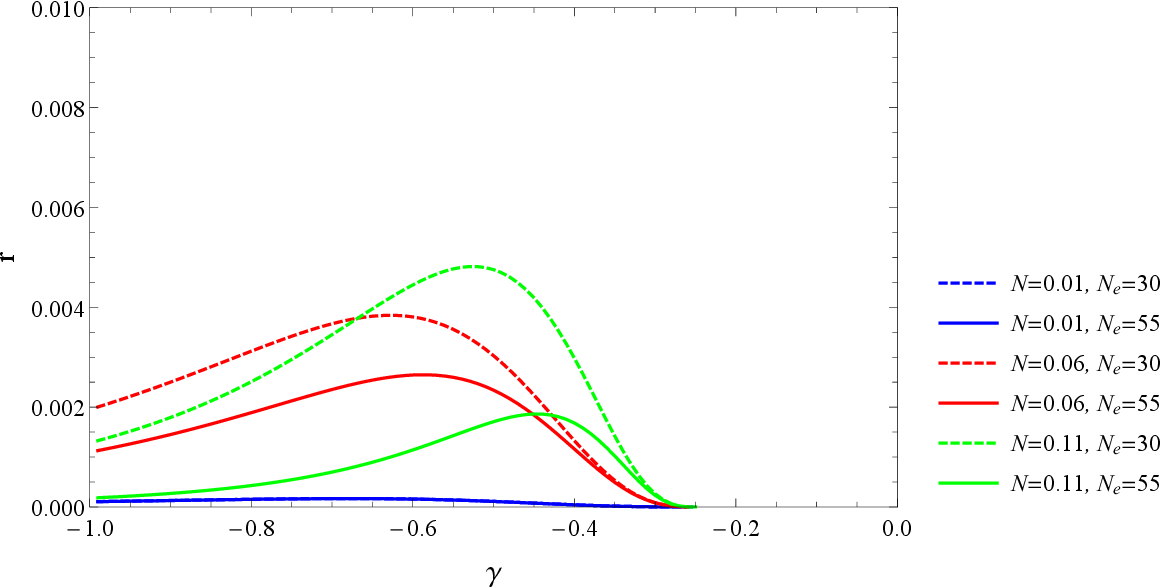} \
\includegraphics[width=0.45\textwidth]{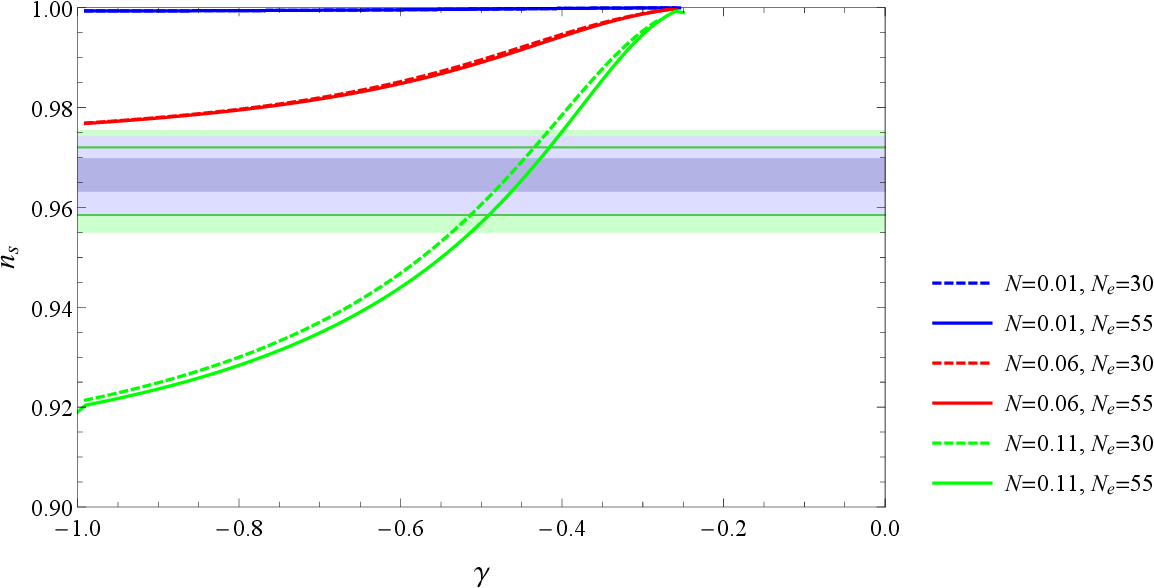} \\\vspace{-0.07cm}
\includegraphics[width=0.45\textwidth]{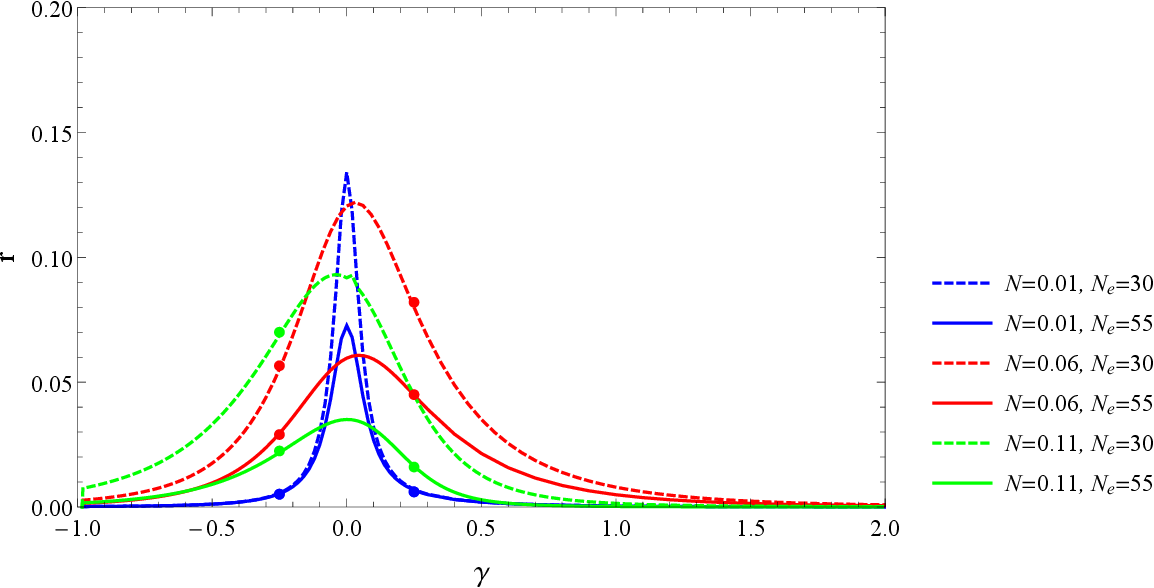}
\includegraphics[width=0.45\textwidth]{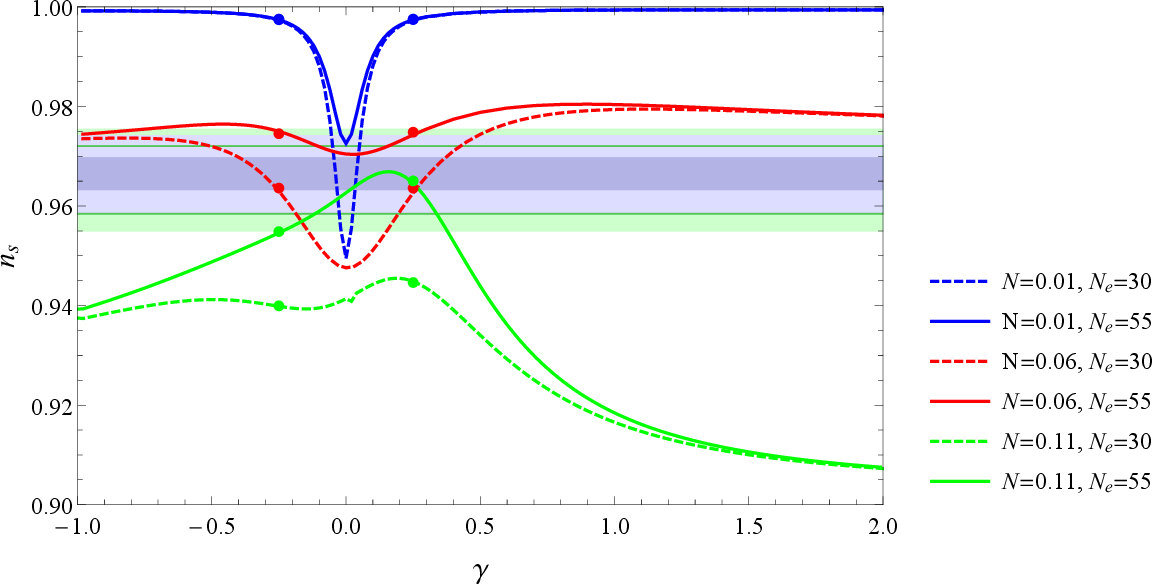}\\\vspace{-0.07cm}
\includegraphics[width=0.45\textwidth]{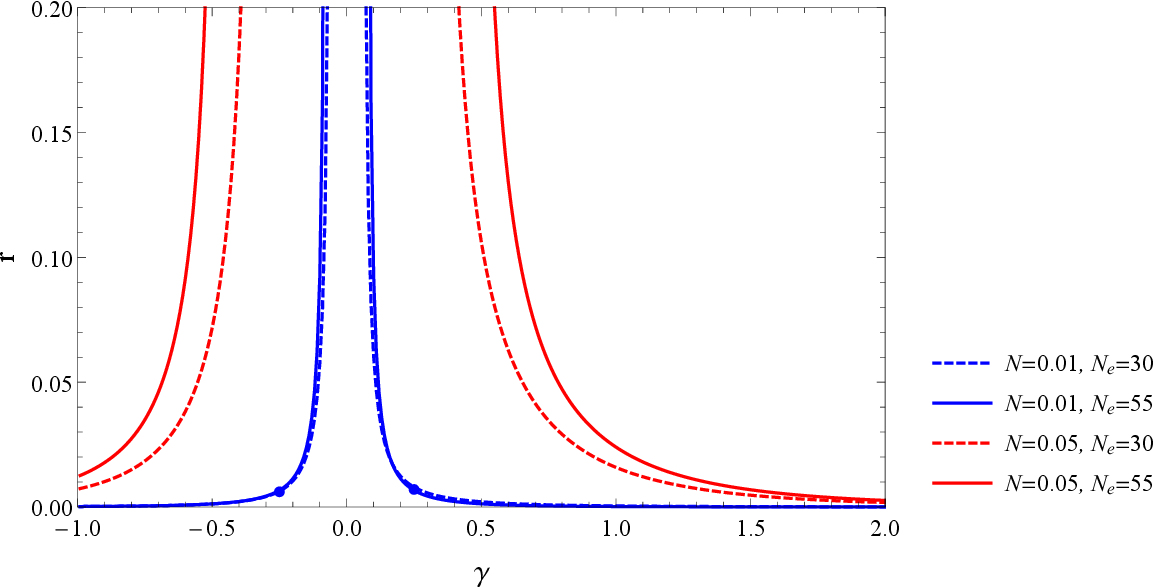}
\includegraphics[width=0.45\textwidth]{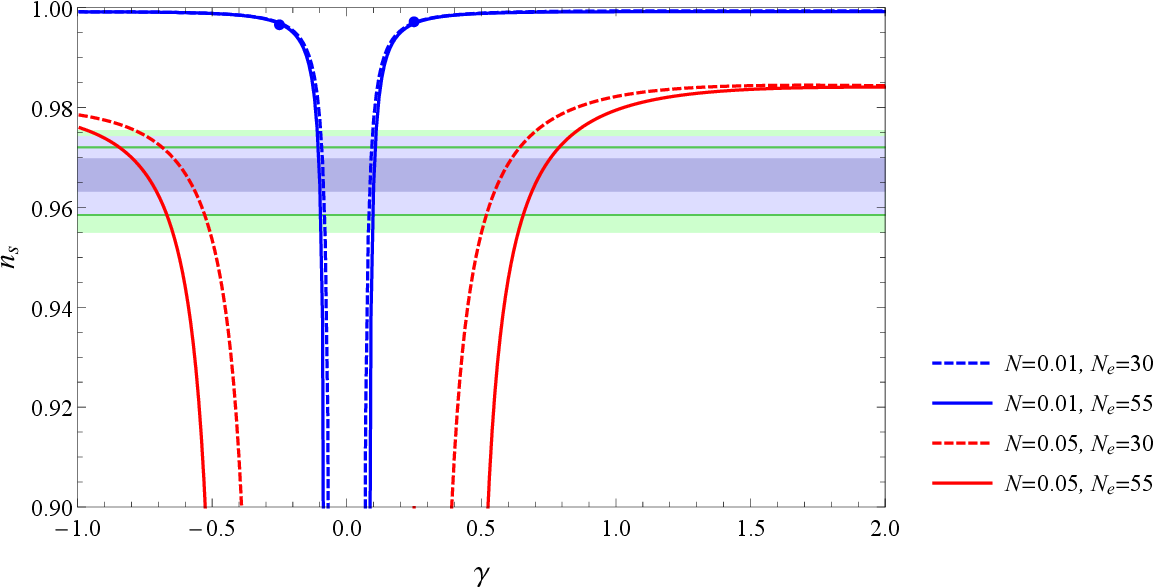} \\\vspace{-0.07cm}
    \includegraphics[width=0.45\textwidth]{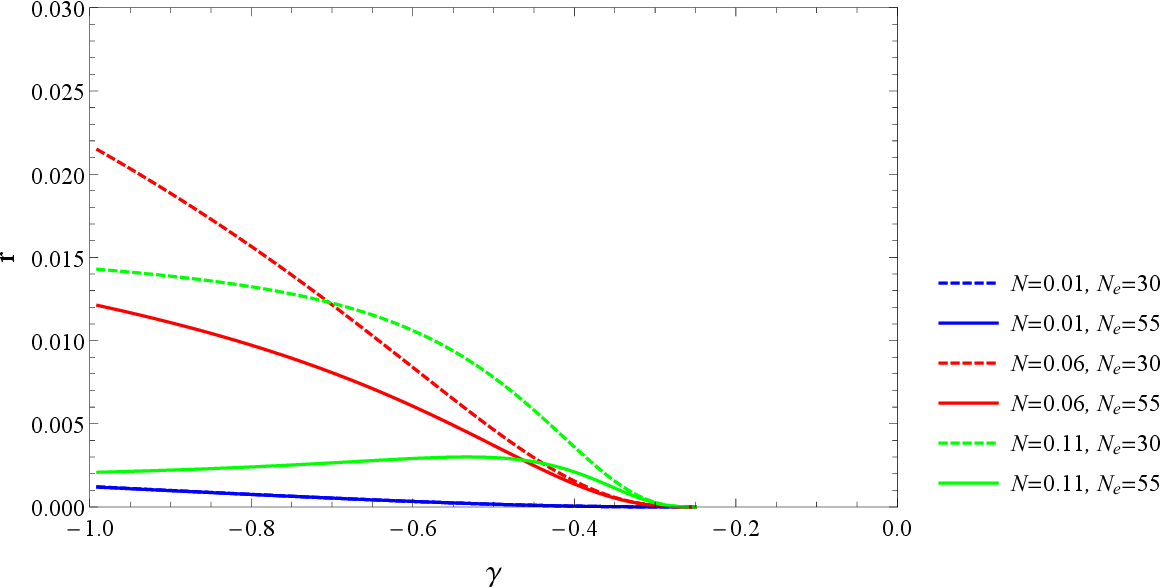}
    \includegraphics[width=0.45\textwidth]{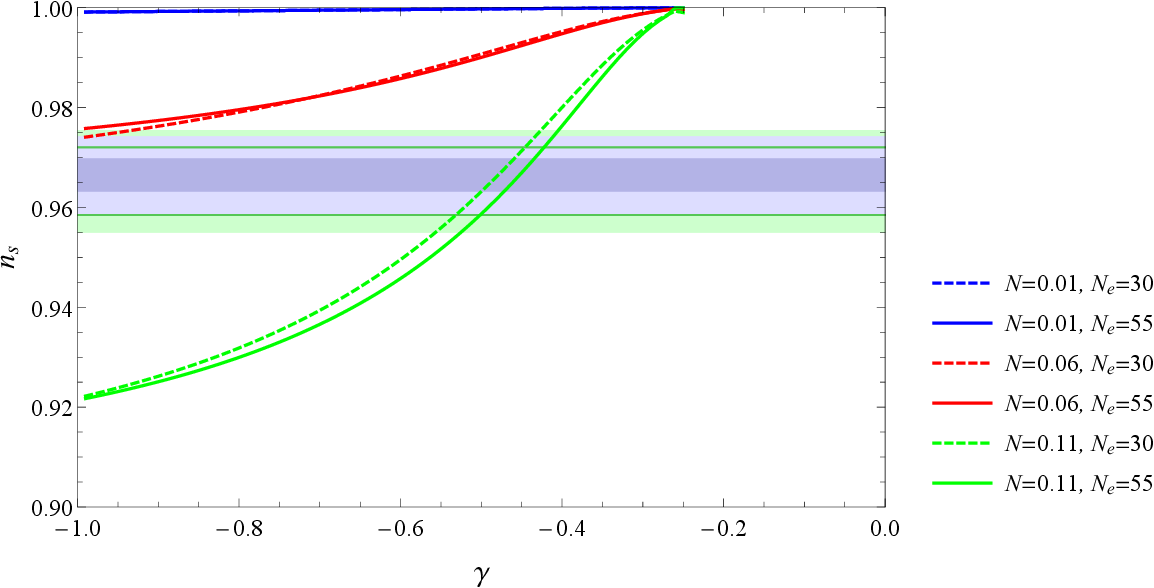}\\\vspace{-0.07cm}

    \includegraphics[width=0.45\textwidth]{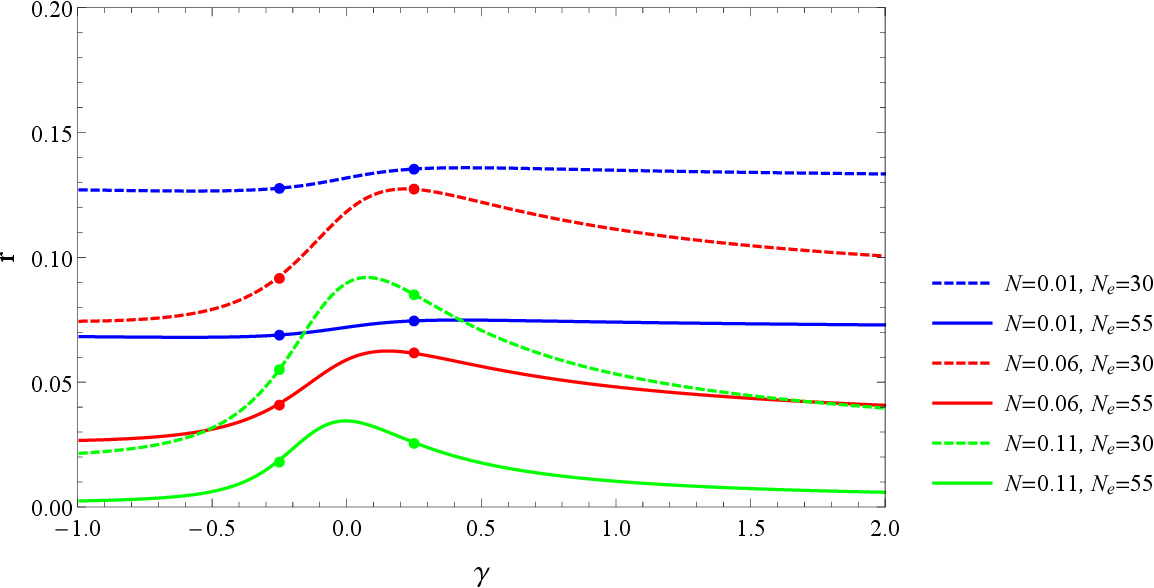}\
    \includegraphics[width=0.45\textwidth]{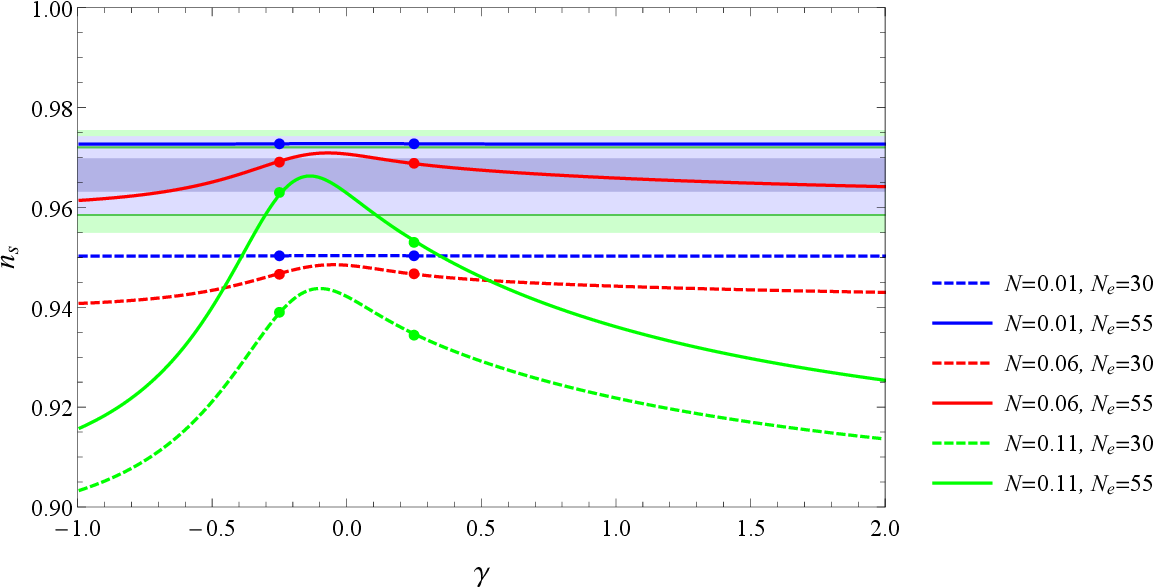} \\\vspace{-0.07cm}
    \caption{ Tensor-to-scalar ratio $r$ and the spectral index $n_s$ as functions of $\gamma$ for the DSG potential (\textcolor{blue}{2}), five cases
        presented in Table \ref{tab1}, $\alpha>0$, several values of $N$, as well as $N_e=30$ and $N_e=55$. In both columns, the highest sub-figures correspond to case $1$, while the lowest sub-figures correspond to case $5$. As mentioned earlier, for $N>0.0547$, case $3$ has no solution for $N_e$ from $30$ to $55$.
        Therefore, in this case, the curves of $r(\gamma)$ and $n_s(\gamma)$ are generated for $N=0.01$ and $N=0.05$. The green and blue regions represent the
        permissible ranges of $n_s$ based on the Planck $2018$ data \cite{Akrami et al. (2020)} and
        the Planck $2018$ data+BK$18$+BAO \cite{Ade et al. (2021)},
        respectively. The $68\%$ regions are highlighted compared to the
        $95\%$ regions \cite{Akrami et al. (2020), Ade et al. (2021)}. In addition,
        bold points representing $\gamma=\pm 0.25$
        in accordance with Eq.~(\ref{e14}), indicating the infinity of
        $N_e$, are hence invalid. As it is obvious from this figure, solutions can be produced for $\gamma>-1$ in cases $2, 3$, and
        $5$, while for cases $1$ and $4$, the range $-1<\gamma<-0.25$ is
        acceptable. In addition, the rate of change in $r$ and $n_s$ increases
        with $\gamma$, as $N$ increases. Moreover, as $\gamma$ increases, $r(\gamma)$ and
        $n_s(\gamma)$ increasingly approach constant values.  The study  also lists
        intervals of $\gamma$ that are compatible with the $95\%$ CL of the Planck $2018$ data \cite {Akrami et al. (2020)} in Table \ref{tab5}.}\label{fig4}
\end{figure*}

Here, we are going to extending the study of the solutions of Eq.~(\ref{e14}), presented in the previous subsection where $\gamma=-0.5$ and $N$ was variable, to the situations in which $\gamma$ is allowed to vary. Our analysis reveals that the acceptable solutions are found for cases $2$, $3$, and $5$ when $\gamma > -1$ (except for $\gamma = 0.25$, which leads to an infinite value of $N_e$), while for cases $1$ and $4$, the valid range is limited to $-1 < \gamma < -0.25$. $\gamma>-0.25$ in cases $1$ and $4$ results in negative potential or imaginary solutions to Eq.~(\ref{e14}), rendering these solutions unacceptable. Moreover, for cases $1-5$, $\gamma \le -1$ causes the same issues and makes the solutions invalid. The admissibility of the range $-1 < \gamma < -0.25$ for all cases led us to choose $\gamma =- 0.5$ in the previous subsection. These intervals for $\gamma$ are further restricted by observational constraints corresponding to specific values of $N$ and $N_e$. To illustrate this, $r(\gamma)$ and $n_s(\gamma)$ are depicted in Fig. \ref{fig4}. To determine the range of $\gamma$ consistent with the particular observational data, represented  by the specific contour in Fig. \ref{fig3}, one identifies the points where the $r(n_s)$ curve (associated with one of cases $1–5$ and the fixed values of $N_e$ and $N$) intersects that contour. For example, the ranges of $\gamma$ that agree with the $95\%$ CL of the Planck $2018$ data \cite{Akrami et al. (2020)} are listed in Table \ref{tab5} (the details of this table, including reheating outcomes, are discussed in Sec. \ref{sec4}). Another important point is that the rate of change in $r$ and $n_s$ increases with $\gamma$ by increasing $N$. This is because for small values of $N$ (such as $N=0.01$) $\varepsilon _{V} $ and $\eta_{V}$, and consequently $r$, $n_s$, and $N_e$ are almost independent of $\gamma$. Moreover,
as $\gamma$ increases, $r(\gamma)$ and $n_s(\gamma)$ gradually
converge to constant values.

\section{Trans-Planckian censorship conjecture}\label{sec3}

The inflationary trans-Planckian condition is arisen due to a
strange expectation for obtaining the sub-Planckian quantum
fluctuations as a direct result of inflationary cosmology in which
such fluctuations can become classical and freeze when their
scales exceed $1/H$
\cite{Martin:2000xs,Brandenberger:2000wr,Brandenberger:2012aj,Kaloper:2002cs,Easther:2002xe}.
Attempts to prevent the presence of such scales as the production
of inflationary cosmology lead to the Trans-Planckian censorship
conjecture (TCC) \cite{Bedroya:2019snp}. Briefly, it states that
the story should progress in such a way that no trans-Planckian
wavelength becomes larger than the supper-Hubble scale ($1/H$) and
thus mathematically, inflation is dealing with the limitation
\cite{Bedroya:2019snp,Lin:2019pmj,Bedroya:2019tba,Brandenberger:2019eni}

\begin{equation}\label{e18}
\frac{a_\mathrm{f}}{a_\mathrm{i}}l_{\mathrm{Pl}}<{\frac{1}{H_\mathrm{f}}},
\end{equation}

\noindent where $l_\mathrm{{Pl}}$ is the Planck length and $H_\mathrm{f}$ denotes the
Hubble parameter at the end of inflation, and it is equivalent with

\begin{equation}\label{e19}
 N_e<\ln({\frac{M_\mathrm{{Pl}}}{H_\mathrm{f}}}),
\end{equation}

\noindent  Considering ${\varepsilon}_V=1$  at the end of inflation $H_f^2$ is calculated as
\begin{equation}\label{e20}
 H_\mathrm{f}^2=\frac{1}{2{M^2_\mathrm{{Pl}}}}V(\phi_\mathrm{f}),
\end{equation}

\noindent By substituting Eq. (\ref{e20}) to Eq. (\ref{e19}), an upper bounds for parameters $\alpha$ and $\beta$ are defined as

\begin{equation}\label{e21}
 \alpha<2{M^4_\mathrm{{Pl}}}\frac{ e^{-2 N_e}}{ -\cos(N\phi_\mathrm{f})+\gamma^{} \cos(2N\phi_\mathrm{f})},
\end{equation}

\begin{equation}\label{e22}
 \beta<2{M^4_\mathrm{{Pl}}}\frac{\gamma e^{-2 N_e}}{ -\cos(N\phi_\mathrm{f})+\gamma^{} \cos(2N\phi_\mathrm{f})}.
\end{equation}

\begin{table}
    \centering \caption{Strictest bound on $\alpha$ under the TCC constraint (\textcolor{blue}{\ref{e19}}) for $\gamma=-0.5$ and cases $1-5$. Assuming a constant $\gamma$, the upper limit of $N$ corresponding to $N_e=55$ in each case, as given in Table \ref{tab4}, provides the strictest bound on $\alpha$. The DSG model can satisfy the TCC constraint in the interval $30 \le N_e \le 55$ for $\alpha$ values below the strictest bound.} \label{tab3}
    \renewcommand{\arraystretch}{1.5}
    \begin{tabular}{|c|c|}
        \hline
        Case & Strictest bound on $\alpha$ (GeV$^4$) \\ \hline
        1 & 9.867$\times 10^{26}$\\
        2 & 1.155$\times 10^{27}$\\
        3 & 2.546$\times 10^{27}$\\
        4 & 9.721$\times 10^{26}$\\
        5 & 1.292$\times 10^{27}$\\ \hline
    \end{tabular}
    
\end{table}

\noindent  Eqs. (\ref{e21}) and (\ref{e22}) show that the parameters $N_e$, $\gamma$, and $N$ determine the strictest bounds on $\alpha$ and $\beta$ under the TCC constraint.  According to Eq. (\ref{e15}), $\phi_\mathrm{f}$ depends on both $N$ and $\gamma$, meaning that its effect is accounted for through these parameters. Among these three parameters, $N_e$ has a particularly strong influence because the exponential term in Eqs. (\ref{e21}) and (\ref{e22}) decreases rapidly with increasing $N_e$. Additionally, $\gamma$ also carries a considerable role. Indeed, enhancement in $-\cos(N\phi_\mathrm{f}) + \gamma \cos(2N\phi_\mathrm{f})$ happens provided that the value of $\gamma$ decreases from $0$ to $-1$ or become greater than $0$. As a result, one is dealing with a decrease in the values of upper bound on $\alpha$. In the limit as $\gamma$ tends to infinity, Eq. (\ref{e21}) reduces to $\alpha < 2 M^4_\mathrm{Pl} \frac{e^{-2 N_e}}{\gamma \cos(2N\phi_\mathrm{f})}$ and Eq. (\ref{e22}) reduces to $\beta < 2 M^4_\mathrm{Pl} \frac{e^{-2 N_e}}{\cos(2N\phi_\mathrm{f})}$, indicating that the strictest bound on $\alpha$ approaches zero, whereas the strictest bound on $\beta$ remains finite. Although $N$ affects the bounds on $\alpha$ and $\beta$, its impact is generally less significant compared to $N_e$ and $\gamma$, since it appears within bounded cosine terms whose values are limited to the interval $[-1, 1]$. Nevertheless, increasing $N$ results in a decrease in the strictest bounds on $\alpha$ and $\beta$. It should be noted that for cases $1$ and $4$, where the allowed range of $\gamma$ is $-1 < \gamma < -0.25$, the strictest bounds on $\alpha$ and $\beta$ remain finite. However, for the other cases, where the permitted range of $\gamma$ is $\gamma > -1$ (except for $\gamma = 0.25$, which leads to divergence of $N_e$), the strictest bound on $\alpha$ tends to zero, as $\gamma$ approaches infinity. In contrast, the strictest bound on $\beta$ remains finite. Based on the above discussion, it can be concluded that for $\gamma = -0.5$ and a chosen case among cases $1$–$5$, the maximum value of the range of $N$ corresponding to $N_e = 55$, consistent with the Planck $2018$ data \cite{Akrami et al. (2020)}, provides the tightest constraint on $\alpha$. For instance, as shown in Table \ref{tab4}, in case $1$ with $N_e = 55$, the viable interval for $N$ is $0.086 \leq N \leq 0.1122$. To determine the strictest bound on $\alpha$, the maximum value $N = 0.1122$ is employed. Table \ref{tab3} presents the strictest bounds on $\alpha$ for $\gamma = -0.5$ and cases $1$–$5$, which are approximately on the order of $10^{27}~\mathrm{GeV}^4$. For values of $\alpha$ below the strictest bound in each case, the DSG model can satisfy the TCC constraint in the interval $30 \le N_e \le 55$.

It should also be noted that since during
inflation, we have $\varepsilon_V,\eta_{V}\ll1$, the Swampland
conditions, expressed as $\frac{\Delta\phi}{M_{Pl}}<c_1$ and
$M_{Pl}\frac{V^\prime(\phi)}{V(\phi)}>c_2$ where $c_i$ are
constants of order one \cite{Mandal:2023zhw}, are not satisfied.
It is due to the fact that de-Sitter space-time is not compatible
with the true quantum gravity theory. Of course, it still deserves
to be studied at the classical level, and more precisely, because
a model with the de-Sitter solution is regarded as a consistent
(low-energy effective) theory
\cite{Ooguri:2006in,Obied:2018sgi,Ooguri:2018wrx,Cicoli:2018kdo,Palti:2019pca,Mandal:2023zhw,Kolb:2021nob}.

\section{Reheating}\label{sec4}

In Sec. \ref{sec2}, for fixed values of $N_e$ and cases $1–5$, the intervals of $N$ and $\gamma$ that are compatible with the Planck $2018$ data \cite{Akrami et al. (2020)} are provided in Tables \ref{tab4} and \ref{tab5}. The section extends this analysis to the reheating phase, with the goal of determining whether these parameter ranges are also responsible for a viable reheating scenario, which is characterized by $N_{\mathrm{reh}}>0$ and $T_{\mathrm{reh}}$ from $10^{-2} \mathrm{GeV}$ to $10^{16} \mathrm{GeV}$ within at least part of the interval $-\frac{1}{3} \leq \omega_{\mathrm{reh}} < \frac{1}{3}$ (the range of $\omega_{\mathrm{reh}}>\frac{1}{3}$ refers to ultra-light particles faster than light and they have not been observed until now. Therefore, this range is not investigated \cite{Nemiroff and Patla (2008)}) \cite{Dai et al. (2014), Mirtalebian et al. (2021), Cheong et al. (2022), Zhou et al. (2022), Afshar et al. (2023)}. As demonstrated in detail in Appendix \textcolor{blue}{I}, $N_{\mathrm{reh}}$ and $T_{\mathrm{reh}}$ are calculated as

\begin{eqnarray}\label{e23}
   && N_{\mathrm{reh}}=\frac{4}{1-3\omega_{\mathrm{reh}}}\Big[ 56.74 \\ &&- \frac{1}{4}\ln \big(\frac{1}{r} \frac{-\cos (N {\phi_\mathrm{f}}) + \gamma \cos (2 N {\phi_\mathrm{f}})}{-\cos (N {\phi_\mathrm{k}})+\gamma \cos (2 N {\phi_\mathrm{k}})}\big)-{N_e}\Big],\nonumber
\end{eqnarray}

\noindent and

\begin{eqnarray}\label{e24}
    &&T_{\mathrm{reh}} = (1.48642\times 10^{16} \text{GeV})\\&&\times\left( r\frac{-\cos (N{\phi_\mathrm{f}})+\gamma \cos (2 N {\phi_\mathrm{f}})}{-\cos (N {\phi_\mathrm{k})+\gamma \cos (2 N {\phi_\mathrm{k}})}} \right)^{0.25}    e^{\frac{-3(1+\omega_{\mathrm{reh}}){N_{\mathrm{reh}}}}{4}}.\nonumber
\end{eqnarray}

\noindent Eq.~(\ref{e23}) indicates that for ${N_e} \ge 56.74$, $N_{\mathrm{reh}}<0$, which is not acceptable. Furthermore, $\omega_{\mathrm{reh}}=\frac{1}{3}$ yields an infinite $N_{\mathrm{reh}}$.  As $N_e$ increases, $N_{\mathrm{reh}}$ decreases while $T_{\mathrm{reh}}$ increases. On the other
hand, by increasing $\omega_{\mathrm{reh}}$, $N_{\mathrm{reh}}$ increases and $T_{\mathrm{reh}}$ decreases. Furthermore, changes in  $N_{\mathrm{reh}}$ and $T_{\mathrm{reh}}$ compared to $N$ and $\gamma$ are very slow and primarily depend on $N_e$ and $\omega_{\mathrm{reh}}$ because of the term

    \begin{table*}
    \centering \caption{Ranges of $N$ that are in agreement with the $95\%$ CL of the Planck  $2018$ TT,TE,EE+lowE+lensing \cite{Akrami et al. (2020)} for $\alpha>0$, $\gamma=-0.5$, five cases listed in Table \ref{tab1}, and $ 30\le N_e\le55$ with a step size of $5$. The intervals of $N_{\mathrm{reh}}$ and $\log_{10}(\frac{T_\mathrm{reh}}{\text{GeV}})$ corresponding to each $N$ range are also presented for $ \omega_{\mathrm{reh}}$ from $-\frac{1}{3}$ to $\frac{1}{6}$ with a step size of $\frac{1}{6}$. As is evident from the table, changes in $N_{\mathrm{reh}}$ and $T_{\mathrm{reh}}$ with respect to $N$  are gradual and mainly controlled by the values of $N_e$ and $\omega_{\mathrm{reh}}$. As $N_e$ increases, $N_{\mathrm{reh}}$ decreases and $T_{\mathrm{reh}}$ increases. Furthermore, when $\omega_{\mathrm{reh}}$ increases, $T_{\mathrm{reh}}$ decreases while $N_{\mathrm{reh}}$ increases. As anticipated from Table \ref{tab6}, for $N_e = 30$, only the case $\omega_{\mathrm{reh}} = -\frac{1}{3}$ satisfies the BBN constraint ($T_{\mathrm{reh}} > 10^{-2}$ GeV). For $N_e = 35$ and $N_e = 40$, both $\omega_{\mathrm{reh}} = -\frac{1}{3}$ and $\omega_{\mathrm{reh}} = -\frac{1}{6}$ are permissible. As $N_e$ increases to $45$ and $50$, the range of acceptable values expands to include $\omega_{\mathrm{reh}} = 0$ as well. Finally, for $N_e = 55$, all four selected values of $\omega_{\mathrm{reh}}$ are valid. The cells that violate the BBN condition are highlighted in red. Yellow-colored cells also denote a scenario where only a subset of the $N$ interval compatible with the Planck $2018$ data \cite{Akrami et al. (2020)} leads to acceptable reheating results. Specifically, for case $1$ and $N_e = 50$, the range $0.105 \leq N \leq 0.1129$ yields $T_{\mathrm{reh}} > 10^{-2}$ GeV, corresponding to $45.11 \leq N_{\mathrm{reh}} \leq 45.55$.} \label{tab4}
    \scriptsize
    \renewcommand{\arraystretch}{1.3}

    \begin{tabular}{p{0.4cm}p{1.8cm}p{1.68cm}p{1.5cm}p{0.01cm}p{1.7cm}p{1.78cm}p{0.01cm}p{1.78cm}p{2cm}p{0.01cm}p{1.8cm}p{1.9cm}}
        \hline\hline
        $ N_e $&  Interval of $N$ &\multicolumn{3}{p{3.3cm}}{$ \ \ \ \ \omega_\mathrm{reh} = -\frac{1}{3} $} &\multicolumn{3}{p{3.3cm}}{$\ \ \ \ \omega_\mathrm{reh} = -\frac{1}{6} $}&\multicolumn{3}{p{3.8cm}} {$ \ \ \ \ \omega_\mathrm{reh} = 0 $}&\multicolumn{2}{p{3.3cm}}{$\ \ \ \ \omega_\mathrm{reh} = \frac{1}{6} $}\vspace{0.01cm}\\
        \cline{3-4}\cline{6-7}\cline{9-10}\cline{12-13}
        &&&&&&&&&&&&\\
        &&$N_\mathrm{reh}$& ${\tiny\log_{10}(\frac{T_\mathrm{reh}}{\text{GeV}})} $ &&$ N_\mathrm{reh} $& $ {\tiny \log_{10}(\frac{T_\mathrm{reh}}{\text{GeV}})} $ &&$ N_\mathrm{reh} $& $ {\tiny \log_{10}(\frac{T_\mathrm{reh}}{\text{GeV}})} $ &&
        $ N_\mathrm{reh} $ & $ {\tiny \log_{10}(\frac{T_\mathrm{reh}}{\text{GeV}})} $ \\
        &&&&&&&&&&&&\\\hline
        30&[0.0890,0.1165]&[51.68,51.80]&[4.12,4.17]&&\cellcolor{red!40}[68.91,69.06]&\cellcolor{red!40}[-3.37,-3.30]&&
        \cellcolor{red!40}[103.36,103.6]&\cellcolor{red!40}[-18.37,-18.27]&&\cellcolor{red!40}[206.73,207.20]&\cellcolor{red!40}[-63.36,-63.16]\\
        35&[0.0883,0.1154]&[41.57,41.74]&[6.28,6.34]&&[55.42,55.65]&[0.24,0.32]&&
        \cellcolor{red!40}[83.14,83.48]&\cellcolor{red!40}[-11.84,-11.71]&&\cellcolor{red!40}[166.28,166.97]&\cellcolor{red!40}[-48.09,-47.81]\\
        40&[0.0876,0.1145]&[31.48,31.68]&[8.45,8.51]&&[41.97,42.25]&[3.87,3.95]&&
        \cellcolor{red!40}[62.96,63.37]&\cellcolor{red!40}[-5.30,-5.15]&&\cellcolor{red!40}[125.92,126.75]&\cellcolor{red!40} [-32.82,-32.50]\\
        45&[0.0871,0.1136]&[21.37,21.63]&[10.62,10.68]&&[28.50,28.84]&[7.49,7.58]&&
        [42.75,43.26]&[1.23,1.39]&&\cellcolor{red!40}[85.51,86.52]&\cellcolor{red!40}[-17.55,-17.17]\\
        50&[0.0865,0.1129]&[11.27,11.57]&[12.79,12.85]&&[15.03,15.43]&[11.12,11.21]&&
        [22.55,23.15]&[7.77,7.95]&&\cellcolor{yellow!40} [45.11,46.30]&\cellcolor{yellow!40} [-2.28,-1.84]\\
        55&[0.0860,0.1122]&[1.17,1.52]&[14.96,15.02]&&[1.57,2.02]&[14.75,14.85]&&
        [2.35,3.04]&[14.30,14.50]&&[4.71,6.08]&[12.98,13.48]\\ \hline
        \multicolumn{13}{c}{Case 1}\\ \hline
        30&[0.0549,0.0829]&[52.48,52.63]&[4.03,4.11]&&\cellcolor{red!40}[69.98,70.17]&\cellcolor{red!40}[-3.57,-3.48]&&
        \cellcolor{red!40}[104.97,105.26]&\cellcolor{red!40}[-18.81,-18.68]&&\cellcolor{red!40}[209.94,210.52]&\cellcolor{red!40}[-64.53,-64.26]\\
        35&[0.0563,0.0860]&[42.35,42.57]&[6.21,6.28]&&[56.47,56.76]&[0.05,0.15]&&
        \cellcolor{red!40}[84.71,85.15]&\cellcolor{red!40}[-12.27,-12.10]&&\cellcolor{red!40}[169.42,170.30]&\cellcolor{red!40}[-49.25,-48.89]\\
        40&[0.0577,0.0888]&[32.21,32.51]&[8.38,8.46]&&[42.95,43.35]&[3.68,3.80]&&
        \cellcolor{red!40}[64.43,65.03]&\cellcolor{red!40}[-5.73,-5.52]&&\cellcolor{red!40}[128.87,130.06]&\cellcolor{red!40}[-33.97,-33.51]\\
        45&[0.0591,0.0911]&[22.07,22.45]&[10.56,10.64]&&[29.43,29.93]&[7.31,7.44]&&
        [44.15,44.90]&[0.80,1.05]&&\cellcolor{red!40}[88.31,89.81]&\cellcolor{red!40}[-18.69,-18.12]\\
        50&[0.0604,0.0932]&[11.93,12.38]&[12.73,12.81]&&[15.91,16.51]&[10.94,11.08]&&
        [23.87,24.77]&[7.35,7.63]&&\cellcolor{red!40}[47.75,49.55]&\cellcolor{red!40}[-3.40,-2.73]\\
        55&[0.0617,0.0949]&[1.79,2.32]&[14.90,14.98]&&[2.39,3.09]&[14.57,14.72]&&
        [3.59,4.64]&[13.90,14.72]&&[7.18,9.28]&[11.88,12.64]\\ \hline
        \multicolumn{13}{c}{Case 2}\\ \hline
        30&[0.0405,0.0485]&[53.03,53.23]&[4.01,4.06]&&\cellcolor{red!40}[70.70,70.97]&\cellcolor{red!40}[-3.66,-3.63]&&
        \cellcolor{red!40}[106.06,106.46]&\cellcolor{red!40}[-19.01,-19.05]&&\cellcolor{red!40}[212.12,212.93]&\cellcolor{red!40}[-65.28,-65.07]\\
        35&[0.0397,0.0469]&[43.04,43.25]&[6.18,6.23]&&[57.39,57.67]&[-0.04,-0.02]&&
        \cellcolor{red!40}[86.09,86.50]&\cellcolor{red!40}[-12.54,-12.50]&&\cellcolor{red!40}[172.19,173.01]&\cellcolor{red!40}[-50.11,-49.89]\\
        40&[0.0388,0.0454]&[33.06,33.27]&[8.35,8.40]&&[44.08,44.36]&[3.56,3.58]&&
        \cellcolor{red!40}[66.12,66.54]&\cellcolor{red!40}[-6.04,-6.00]&&\cellcolor{red!40}[132.25,133.08]&\cellcolor{red!40}[-34.93,-34.72]\\
        45&[0.0380,0.0441]&[23.08,23.28]&[10.52,10.57]&&[30.77,31.04]&[7.18,7.20]&&
        [46.16,46.57]&[0.46,0.50]&&\cellcolor{red!40}[92.32,93.14]&\cellcolor{red!40}[-19.76,-19.54]\\
        50&[0.0373,0.0429]&[13.09,13.30]&[12.69,12.74]&&[17.45,17.73]&[10.80,10.82]&&
        [26.18,26.60]&[6.96,7.01]&&\cellcolor{red!40}[53.20,52.37]&\cellcolor{red!40}[-4.58,-4.36]\\
        55&[0.0366,0.0418]&[3.10,3.31]&[14.86,14.91]&&[4.14,4.41]&[14.41,14.43]&&
        [6.21,6.62]&[13.47,13.51]&&[12.42,13.24]&[10.60,10.81]\\ \hline
        \multicolumn{13}{c}{Case 3}\\ \hline
        30&[0.0924,0.1195]&[51.91,52.02]&[4.13,4.17]&&\cellcolor{red!40}[69.21,69.36]&\cellcolor{red!40}[-3.39,-3.33]&&
        \cellcolor{red!40}[103.82,104.04]&\cellcolor{red!40}[-18.46,-18.36]&&\cellcolor{red!40}[207.64,208.08]&\cellcolor{red!40}[-63.64,-63.45]\\
        35&[0.0915,0.1181]&[41.81,41.96]&[6.29,6.34]&&[55.75,55.95]&[0.22,0.29]&&
        \cellcolor{red!40}[83.62,83.93]&\cellcolor{red!40}[-11.92,-11.81]&&\cellcolor{red!40}[167.25,167.86]&\cellcolor{red!40}[-48.37,-48.12]\\
        40&[0.0906,0.1169]&[31.71,31.91]&[8.46,8.51]&&[42.28,42.55]&[3.84,3.92]&&
        \cellcolor{red!40}[63.43,63.82]&\cellcolor{red!40}[-5.39,-5.25]&&\cellcolor{red!40}[126.86,127.65]&\cellcolor{red!40}[-33.10,-32.80]\\
        45&[0.0898,0.1158]&[21.61,21.85]&[10.63,10.68]&&[28.82,29.13]&[7.47,7.55]&&
        [43.23,43.70]&[1.14,1.29]&&\cellcolor{red!40}[86.47,87.41]&\cellcolor{red!40}[-17.83,-17.48]\\
        50&[0.0890,0.1148]&[11.52,11.80]&[12.80,12.85]&&[15.36,15.74]&[11.09,11.18]&&
        [23.04,23.61]&[7.67,7.85]&&\cellcolor{red!40}[46.08,47.22]&\cellcolor{red!40}[-2.57,-2.15]\\
        55&[0.0884,0.1140]&[1.42,1.75]&[14.97,15.02]&&[1.89,2.33]&[14.72,14.81]&&
        [2.84,3.50]&[14.21,14.40]&&[5.69,7.00]&[12.69,13.16]\\ \hline
        \multicolumn{13}{c}{Case 4}\\ \hline
        40&(0,0.0495]&[33.22,33.59]&[8.35,8.41]&&[44.29,44.79]&[3.48,3.60]&&
        \cellcolor{red!40}[66.44,67.19]&\cellcolor{red!40}[-6.23,-6.01]&&
        \cellcolor{red!40}[132.88,134.38]&\cellcolor{red!40}[-35.41,-34.87]\\
        45&(0,0.0694]&[22.88,23.56]&[10.50,10.60]&&[30.50,31.42]&[7.09,7.29]&&
        [45.76,47.13]&[0.27,0.67]&&\cellcolor{red!40}[91.52,94.26]&\cellcolor{red!40}[-20.19,-19.20]\\
        50&(0,0.0785]&[12.63,13.54]&[12.66,12.79]&&[16.85,18.05]&[10.70,10.96]&&
        [25.27,27.08]&[6.78,7.30]&&\cellcolor{red!40}[50.55,54.16]&
        \cellcolor{red!40}[-4.97,-3.67]\\
        55&(0,0.0843]&[2.42,3.51]&[14.83,14.97]&&[3.23,4.68]&[14.32,14.62]&&
        [4.85,7.03]&[13.30,13.91]&&[9.71,14.06]&[10.24,11.80]\\ \hline
        \multicolumn{13}{c}{Case 5}\\ \hline
    \end{tabular}
   
\end{table*}

\begin{eqnarray}
    \text{``}\frac{1}{4}\ln \big( \frac{1}{r} \frac{-\cos (N \phi_\mathrm{f}) + \gamma \cos (2 N \phi_\mathrm{f})}{-\cos (N {\phi_\mathrm{k}})+\gamma \cos (2 N {\phi_\mathrm{k}})}\big)\text{"}\nonumber
\end{eqnarray}
in Eq.~(\ref{e23}) and the term
\begin{eqnarray}
    \text{``}\left(r \frac{-\cos (N \phi_\mathrm{f})+\gamma \cos (2 N \phi_\mathrm{f})}{-\cos (N \phi_\mathrm{k})+\gamma \cos (2 N {\phi_\mathrm{k}})}\right)^{0.25}\text{"}\nonumber
\end{eqnarray}
 in Eq.~(\ref{e24}) are significantly smaller than the other terms and coefficients. Therefore, the reheating final temperature can approximately be written as

\begin{equation}\label{e25}
\ln(\frac{T_{\mathrm{reh}}}{\text{GeV}}) \approx \ln(1.48642\times 10^{16})-3 ( \frac{1+\omega_{\mathrm{reh}}}{1-3\omega_{\mathrm{reh}}}) [56.74-N_e].
\end{equation}
 The case $\omega_{\mathrm{reh}} = -\frac{1}{3}$ is special, as the corresponding $T_{\mathrm{reh}}$ is always greater than the lower bound of $10^{-2} \mathrm{GeV}$ for all values of $N_e$ in the range $30$ to $55$. In addition, for each $N_e$, the approximate upper bound on $\omega_{\mathrm{reh}}$ can be derived by setting $T_{\mathrm{reh}} = 10^{-2} \mathrm{GeV}$ in Eq.~(\ref{e25}). The approximate upper bounds on $\omega_{\mathrm{reh}}$ for various values of $N_e$ are listed in Table \ref{tab6}. Accordingly, for $N_e = 30$, the range $-\frac{1}{3} \leq \omega_{\mathrm{reh}} \lesssim -0.187$ yields acceptable results, whereas for $N_e = 55$, the range $-\frac{1}{3} \leq \omega_{\mathrm{reh}} \lesssim 0.28$ is permissible. It should be mentioned that Eqs.~(\ref{e23}) and (\ref{e24}) can be used to determine the precise bound for a given case from cases $1–5$ and specific values of $N$, $\gamma$, and $N_e$.

\begin{table*}
    \centering\caption{Ranges of $\gamma$ that are in agreement with the $95\%$ CL of the Planck  $2018$ TT,TE,EE+lowE+lensing \cite{Akrami et al. (2020)} for $\alpha>0$, five cases listed in Table \ref{tab1}, several values of $N$, as well as $N_e=30$ and $N_e=55$. The intervals of $N_{\mathrm{reh}}$ and $\log_{10}(\frac{T_\mathrm{reh}}{\text{GeV}})$ corresponding to each $\gamma$ range are presented for $ \omega_{\mathrm{reh}}$ from $-\frac{1}{3}$ to $\frac{1}{6}$ with a step size of $\frac{1}{6}$. As is evident from the table, changes in $N_{\mathrm{reh}}$ and $T_{\mathrm{reh}}$ with respect to $\gamma$ are gradual and mainly controlled by the values of $N_e$ and $\omega_{\mathrm{reh}}$. As $N_e$ increases, $N_{\mathrm{reh}}$ decreases and $T_{\mathrm{reh}}$ increases. Furthermore, when $\omega_{\mathrm{reh}}$ increases, $T_{\mathrm{reh}}$ decreases while $N_{\mathrm{reh}}$ increases. As anticipated from Table \ref{tab6}, for $N_e = 30$, only the cells corresponding to $\omega_{\mathrm{reh}} = -\frac{1}{3}$ are acceptable, whereas for $N_e = 55$, all four selected values of $\omega_{\mathrm{reh}}$ are permissible. The cells that violate the BBN condition are highlighted in red.}\label{tab5}
    \scriptsize
    \renewcommand{\arraystretch}{1.3}
    \begin{tabular}{p{0.4cm}p{0.5cm}p{1.71cm}p{1.65cm}p{1.68cm}p{0.001cm}p{1.68cm}p{1.75cm}p{0.001cm}p{1.78cm}p{2cm}p{0.001cm}p{1.8cm}p{1.75cm}}
        \hline
        \hline
        $ N_e $ & $N$ & Interval of $\gamma$ &\multicolumn{3}{p{3.2cm}}{$ \ \ \ \ \omega_{\mathrm{reh}} = -\frac{1}{3} $} &\multicolumn{3}{p{3.2cm}}{$\ \ \ \ \omega_{\mathrm{reh}} = -\frac{1}{6} $}&\multicolumn{3}{p{3.3cm}} {$ \ \ \ \ \omega_{\mathrm{reh}} = 0 $}&\multicolumn{2}{p{3.3cm}}{$\ \ \ \ \omega_{\mathrm{reh}} = \frac{1}{6} $}\vspace{0.01cm}\\
        \cline{4-5}\cline{7-8}\cline{10-11}\cline{13-14}
        &&&&&&&&&&&&&\\
        &   &&$N_{\mathrm{reh}}$& ${\tiny\log_{10}(\frac{T_{\mathrm{reh}}}{\text{GeV}})} $ &&$ N_{\mathrm{reh}} $& $ {\tiny \log_{10}(\frac{T_{\mathrm{reh}}}{\text{GeV}})} $ &&$ N_{\mathrm{reh}} $& $ {\tiny \log_{10}(\frac{T_{\mathrm{reh}}}{\text{GeV}})} $ &&
        $ N_{\mathrm{reh}} $ & $ {\tiny \log_{10}(\frac{T_{\mathrm{reh}}}{\text{GeV}})} $ \\
        &&& &&&&&&&&&&\\\hline
        30&0.11&[-0.54,-0.42]&[51.6, 51.72]&[4.15,4.17]&&\cellcolor{red!40}[68.8,68.96]&\cellcolor{red!40}[-3.222,-3.181]&&\cellcolor{red!40}[103.2,103.44]&\cellcolor{red!40}[-18.3,-18.26]&&\cellcolor{red!40}[206.4,206.88]&\cellcolor{red!40}[-63.22,-63.07]\\
        55&0.11&[-0.51,-0.4]&[1.2,1.24]&[15,15.02]&&[1.6,1.65]&[14.82,14.84]&&  [2.4,2.48]&[14.46,15]&&[4.79,4.96]&[13.39,13.46]\\ \hline
        \multicolumn{14}{c}{Case 1}\\ \hline
        30&0.01&[-0.03,0.03]&[53.6,53.61]&[4.012,4.014]&&\cellcolor{red!40}[71.46,71.48]&\cellcolor{red!40}[-3.748,-3.744]&&\cellcolor{red!40}[107.2,107.22]&\cellcolor{red!40}[-19.269,-19.263]&&\cellcolor{red!40}[214.39,214.43]&\cellcolor{red!40}[-65.83,-65.82]\\
        30&0.06&(-1,-0.16]&(51.67,53.37]&[4.05,4.06)&&\cellcolor{red!40}(68.89,71.16]&\cellcolor{red!40}[-3.67,-3.43)&&\cellcolor{red!40}(103.34,106.74]&\cellcolor{red!40}[-19.12,-18.39)&&\cellcolor{red!40}(206.67,213.5]&\cellcolor{red!40}[-65.47,-63.26)\\
        30&0.06&[0.16,0.54]&[53.15,53.6]&[3.947,4.013]&&\cellcolor{red!40}[70.87,71.47]&\cellcolor{red!40}[-3.747,-3.746]&&\cellcolor{red!40}[106.3,107.21]&\cellcolor{red!40}[-19.27,-19.14]&&\cellcolor{red!40}[212.61,214.41]&\cellcolor{red!40}[-65.8,-65.3]\\
        55&0.01&[-0.023,0.023]&[3.49,4]&[14.822,14.824]&&[4.65,4.66]&[14.315,14.319]&&[6.97,6.99]&[13.303,13.309]&&[13.95,13.99]&[9.49,10.27]\\
         55&0.06&(-1,-0.74]&(1.35,1.82] & [14.908,14.9082)&& (1.8,2.43]& [14.65,14.71)&&(2.7,3.64]& [14.12,14.32)&&(5.4,7.28]&[12.54,13.15)\\
        55&0.06&[-0.3,0.3]&[2.82,3.28]&[14.8,14.89]&&[3.75,4.38]&[14.32,14.49]&&[5.63,6.57]&[13.37,13.67]&&[11.26,13.13]&[10.52, 11.22]\\
        55&0.11&[-0.37,0.23]&[2.69,4.47]&[14.8,15]&&[3.53,4.47]&[14.32,14.51]&&[5.3,6.71]&[13.35,13.74]&&[10.59,13.42]&[10.43,11.49]\\ \hline
        \multicolumn{14}{c}{Case 2}\\ \hline
        30&0.01&[-0.096,-0.084]&[53.47,53.59]&[3.97,4.01]&&\cellcolor{red!40}[71.3,71.45]&\cellcolor{red!40}[-3.77,-3.74]&&\cellcolor{red!40}[106.95,107.17]&\cellcolor{red!40}[-19.257,-19.252]&&\cellcolor{red!40}[213.89,214.35]&\cellcolor{red!40}[-65.8,-65.7]\\
        30&0.01&[0.084,0.096]&[53.51,53.61]&[3.96,4]&&\cellcolor{red!40}[71.35,71.49]&\cellcolor{red!40}[-3.78,-3.75]&&\cellcolor{red!40}[107.02,107.23]&\cellcolor{red!40}[-19.277,-19.274]&&\cellcolor{red!40}[214.04,214.45]&\cellcolor{red!40}[-65.84,-65.75]\\
        30&0.05&[-0.79,-0.52]&[52.62,53.22]&[4.03,4.07]&&\cellcolor{red!40}[70.15,70.96]&\cellcolor{red!40}[-3.63,-3.59]&&\cellcolor{red!40}[105.23,106.44]&\cellcolor{red!40}[-19.04,-18.82]&&\cellcolor{red!40}[210.46,212.87]&\cellcolor{red!40}[-65.26,-64.52]\\
        30&0.05&[0.53,0.7]&[53.3,53.57]&[3.95,4]&&\cellcolor{red!40}[71.06,71.43]&\cellcolor{red!40}[-3.77,-3.76]&&\cellcolor{red!40}[106.59,107.14]&\cellcolor{red!40}[-19.27,-19.2]&&\cellcolor{red!40}[213.18,214.29]&\cellcolor{red!40}[-65.8,-65.49]\\
        55&0.01&[-0.108,-0.098]&[3.48,3.59]&[14.83,14.87]&&[4.64,4.78]&[14.33,14.35]&&[6.96,7.18]&[13.315,13.32]&&[13.91,14.35]&[10.2,10.3]\\
        55&0.01&[0.098,0.108]&[3.52,3.62]&[14.82,14.87]&&[4.69,4.83]&[14.31,14.34]&&[7.04,7.24]&[13.293,13.296]&&[14.08,14.48]&[10.15,10.24]\\
        55&0.05&[-0.97,-0.66]&[2.55,3.19]&[14.89,14.93]&&[3.4,4.25]&[14.47,14.52]&&[5.1,5.38]&[13.54,13.78]&&[10.2,12.76]&[10.77,11.56]\\
        55&0.05&[0.66,0.86]&[3.24,3.56]&[14.81,14.86]&&[4.32,4.75]&[14.34,14.343]&&[6.48,7.12]&[13.31,13.4]&&[12.96,14.24]&[10.22,10.59]\\ \hline
        \multicolumn{14}{c}{Case 3}\\ \hline
        30&  0.06 &(-1.-0.932] &[52.64,52.68) & (4.07,4.074] && \cellcolor{red!40}[70.19,70.23)& \cellcolor{red!40}(-3.55,-3.546] && \cellcolor{red!40}[105.28,105.35)&\cellcolor{red!40}(-18.8,-18.79] &&\cellcolor{red!40}[210.56,210.70)&\cellcolor{red!40}(-64.55,-64.51]\\
        30&0.11&[-0.56,-0.43]&[51.76,52.06]&[4.15,4.17]&&\cellcolor{red!40}[69.01,69.41]&\cellcolor{red!40}[-3.37,-3.34]&&\cellcolor{red!40}[103.52,104.12]&\cellcolor{red!40}[-18.44,-18.32]&&\cellcolor{red!40}[207.04,208.24]&\cellcolor{red!40}[-63.66,-63.28]\\
        55&0.11&[-0.52,-0.4]&[1.35,1.49]&[15,15.02]&&[1.8,1.99]&[14.804,14.808]&&[2.7,2.98]&[14.37,14.42]&&[5.4,5.96]&[13.08,13.24]\\ \hline
        \multicolumn{14}{c}{Case 4}\\ \hline
        55&0.01&$(-1,\infty)$&(3.5,4.85)&(14.4,14.83)&&(4.67,6.47)&(13.7,14.33)&&(7,9.7)&(12.29,13.31)&&(14,19.41)&(8.07,10.27)\\
        55&0.06&$(-1,\infty)$&(2.9,4.35)&(14.49,14.91)&&(3.86,5.8)&(13.86,14.49)&&(5.79,8.71)&(12.6,13.64)&&(11.59,17.41)&(8.82,11.13)\\
        55&0.11&[-0.34,0.18]&[4.13,4.88]&[15.17,15.42]&&[5.5,6.51]&[14.58,14.72]&&[8.25,9.76]&[13.3,13.38]&&[16.511,19.52]&[9.07,9.8]\\ \hline
        \multicolumn{14}{c}{Case 5}\\ \hline
    \end{tabular}
\end{table*}

In addition to the ranges of $N$ and $\gamma$ that are consistent with the Planck 2018 data \cite{Akrami et al. (2020)}, Tables \ref{tab4} and \ref{tab5} also present the corresponding ranges of $N_{\text{reh}}$ and $\log_{10}\left( \frac{T_\mathrm{reh}}{\text{GeV}} \right)$ for $\omega_{\mathrm{reh}}$ varying from $-\frac{1}{3}$ to $\frac{1}{6}$ with a step size of $\frac{1}{6}$. The described relationship between $N_{\mathrm{reh}}$ and $T_{\mathrm{reh}}$ with respect to the parameters $N$, $\gamma$, $N_e$, and $\omega_{\mathrm{reh}}$ can be observed in these tables. Aligned with Table \ref{tab6}, these tables show that $N_e = 30$ satisfies the reheating conditions for $\omega_{\mathrm{reh}} = -\frac{1}{3}$. On the other hand, for $N_e = 55$, the conditions are satisfied for all four selected $\omega_{\mathrm{reh}}$. Furthermore, for case $1$ with $\gamma = -0.5$ and $N_e = 50$, the range $0.0865 \leq N \leq 0.1129$ is compatible with the $95\%$ CL of the Planck $2018$ data \cite{Akrami et al. (2020)}. However, for $\omega_{\mathrm{reh}} = \frac{1}{6}$, only a subset of this range leads to appropriate reheating results, such that the sub-interval $0.105 \leq N \leq 0.1129$ produces $T_{\mathrm{reh}} > 10^{-2}$ GeV. In the present scenario, the allowed range for $N_{\mathrm{reh}}$ is $45.11 \leq N_{\mathrm{reh}} \leq 45.55$.

\section{Summary and Discussions}

Despite the many successes of the Big Bang theory \cite{Hubble (1929),Penzias and Wilson (1965)}, it suffers from challenges including the flatness \cite{Dicke (1970),Dicke and Peebles (1979)}, horizon \cite{Rindler (1956)}, and magnetic monopole problems \cite{Dirac (1931)}. These issues are assumed to be solvable (or at least relaxed) by using the inflation theory based on introducing high-energy inflaton fields \cite{Guth (1981),Starobinsky (1980), Sato (1981), Linde (1982), Albrecht and Steinhardt (1982), Hawking et al. (1982), Linde (1983)}.
In this regard, motivated by the properties of Soliton fields \cite{Mukhanov (2005), Hobson et al. (2006), Durrer (2008), Roos (2003), Kibble (1976), Vilenkin and Shellard (1994), Riotto and Trodden (2002)}, various attempts have been made to employ Soliton fields in the study of inflation and the early universe in various setups. However, despite the fact that  the reheating era plays a crucial role in connecting the achievements of inflationary and Big Bang theories \cite{Amin et al. (2015)}, no previous study has addressed reheating for the DSG potential. Moreover, the consistency between solitonic potentials and observational data has only been examined for the PNGBs potential, using previous Planck data \cite{Freese and Kinney (2015), Mielke (2020)}. It is also important to note that respecting the TCC constraint (\ref{e19}) is necessary for inflationary models expected to be closer to the real early universe. In the previous studies, TCC has not been addressed. Here, we investigate the DSG potential as a model for the early universe and evaluate its consistency with the Planck $2018$ data \cite{Akrami et al. (2020)} and the Planck $2018$+BK$18$+BAO data \cite{Ade et al. (2021)}. We also assess its ability to satisfy the TCC constraint (\ref{e19}) and to produce a number of e-foldings in the range $30 \le N_e \le 55$. Furthermore, this paper provides a consistent description of the reheating phase, yielding a positive reheating number of e-foldings ($N_{\text{reh}}$) and a reheating final temperature ($T_{\text{reh}}$) ranging from $10^{-2}$ GeV to $10^{16}$ GeV for $-\frac{1}{3} \leq \omega_{\text{reh}} < \frac{1}{3}$.

According to our analysis, the values of $\alpha < 0$ and/or $\gamma \le -1$ result in a negative potential that violates the slow-roll condition or yield imaginary solutions to Eq.~(\ref{e14}), making them invalid; therefore, only the scenarios where $\alpha > 0$ and $\gamma > -1$ are taken into consideration. Five cases are defined in Table \ref{tab1} that classify the possible sign combinations of $V'(\phi)$ and $1 - 4\gamma\cos(N\phi)$, which are crucial in ensuring acceptable solutions to
Eq. (\ref{e14}). In cases $2$, $3$, and $5$, acceptable values of $r$ and $n_s$ for fixed $N_e$ can be obtained when $\gamma > -1$ (except for $\gamma = 0.25$, which leads to an infinite value of $N_e$). The allowed values of $\gamma$ lies within $-1 < \gamma < -0.25$ for cases $1$ and $4$. If $\gamma > -0.25$, then the resulting potential becomes negative or Eq.~(\ref{e14}) generates imaginary solutions and thus, this range is impermissible. Additionally, the admissible interval for $\gamma$ can be further refined by examining the consistency of $r$ and $n_s$ while keeping $N$ fixed. As an example, for case $1$ with $N_e = 30$ and $N = 0.11$, the allowed range for $\gamma$ is $-0.54 \leq \gamma \leq -0.42$. Furthermore, a range of $N$ can be established for each $\gamma$ and $N_e$ consistent with the Planck $2018$
data \cite {Akrami et al. (2020)} and the Planck $2018$ data+BK$18$+BAO \cite{Ade et al. (2021)}. For instance, agreement with the $95\%$ CL of the Planck $2018$ data \cite {Akrami et al. (2020)} is obtained for $\gamma = -0.5$ and $N_e = 55$ in the interval $0.086 \leq N \leq 0.1122$ for case $1$ and $0 \leq N \leq 0.0843$ for case $5$.

\begin{table}
    \centering \caption{Approximate upper bounds on $\omega_{\mathrm{reh}}$ for $N_e$ from $30$ to $55$ with a step size of $5$ derived from Eq.~(\ref{e25}) for $T_{\mathrm{reh}}=10^{-2} \text{GeV}$. The valid range of $\omega_{\mathrm{reh}}$ extends from $ -\frac{1}{3}$ to the approximate upper bound corresponding to each $N_e$. Eqs.~(\ref{e23}) and (\ref{e24}) provide an exact bound for a given case from cases $1-5$ and selected values of $N$, $\gamma$, and $N_e$.} \label{tab6}
    \renewcommand{\arraystretch}{1.5}
    \begin{tabular}{|c|c|}
        \hline
        $N_e$ & Approximate upper bound on $\omega_\mathrm{reh}$ \\ \hline
        30 & -0.187 \\
        35 & -0.123\\
        40 & -0.048 \\
        45 & 0.041 \\
        50 & 0.148\\
        55& 0.28  \\\hline
    \end{tabular}
    
\end{table}

Satisfying the TCC condition is one of the essential requirements for a successful inflationary model. In this work, we have examined the constraints imposed by the TCC condition (\textcolor{blue}{19}) on the free parameters $\alpha$ and $\beta$. Our analysis shows that in cases $1$ and $4$, where the allowed range of $\gamma$ is $-1 < \gamma < -0.25$, the strictest bounds, as defined by Eqs. (\ref{e21}) and (\ref{e22}), on both $\alpha$ and $\beta$ remain finite. In contrast, for the other cases, where the permitted range is $\gamma > -1$, the strictest bound on $\alpha$ tends to zero as $\gamma$ approaches infinity, while the bound on $\beta$ remains finite. As an example, for $\gamma = -0.5$, the strictest bounds on $\alpha$ in cases $1–5$ are found to be approximately on the order of $10^{27}~\mathrm{GeV}^4$, as shown in Table \ref{tab3}. This implies that if $\alpha$ lies below the corresponding bound in each case, the DSG model satisfies the TCC condition within the interval $30 \le N_e \le 55$. Furthermore, the results presented in Tables \ref{tab4}-\ref{tab6} demonstrate that each $N_e$ from $30$ to $55$ can satisfy the reheating conditions within a portion of the interval $-\frac{1}{3} \leq \omega_{\mathrm{reh}} < \frac{1}{3}$. As example, acceptable values of $\omega_{\mathrm{reh}}$ fall within the range $-\frac{1}{3} \leq \omega_{\mathrm{reh}} \lesssim -0.187$ for $N_e = 30$, while for $N_e = 55$, the interval $-\frac{1}{3} \leq \omega_{\mathrm{reh}} \lesssim 0.28$ is viable.

In general, we are dealing with a multi-parameter model. Indeed, for this model and all models involving multidimensional parameter space, being equipped with methods like Markov Chain Monte Carlo (MCMC) enables us to conduct more detailed studies in the future. Moreover, investigating the consequences of modelling the early universe slow-roll inflationary scenario is crucial; it is just the starting point, and other scenarios such as constant-roll
inflation \cite{Inoue and Yokoyama (2002), Namjoo et al. (2013), Martin et al. (2013), Motohashi et al. (2015)} and warm inflation
\cite{Berera (1995)} are beneficial to be investigated.
Furthermore, examining solitonic potentials within the framework
of modified theories of gravity may yield valuable information
about the early cosmos and the employed theory. Finally, it should
be noted that analysing the solitonic models by using the
gravitational waves' data such as NANOGrav \cite{Vagnozzi (2021),Li (2020), Kuroyanagi (2021), Cai (2021), Li and Shapiro (2021),
Benetti et al. (2022), Ashoorioon et al. (2022), Vagnozzi (2023),
Antoniadis et al. (2023), Borah et al. (2023), Datta (2023), Niu
and Rahat (2023)} helps us obtain more appropriate models and
indeed, has enough potential to be considered as future projects.
In this regard, it is also worth noting that a real inflaton
should have the ability to create bosons. This is the reason for the importance of the preheating stage, a period between inflation and reheating in which bosons are created \cite{Kofman:1994rk, Kofman:1997yn, Bassett:2005xm, Khlebnikov:1996zt, Khlebnikov:1996wr}, which motivates us to focus on this stage in future work.

\section*{Appendix I}

The appendix provides detailed calculations for Eqs. (\ref{e23}) and (\ref{e24}) in the reheating section. In this regard, according to the Bianchi identity
(${G_{\mu\nu}^{;\mu}}=0$), we have

\begin{equation}\label{e26}
    \dot{\rho}+ 3H(\rho+p)=0,
\end{equation}

\noindent combined with $\omega=\frac{p}{\rho}$ to reach at

\begin{equation}\label{e27}
    \rho= \rho_{0}a^{-3(1+\omega)}.
\end{equation}

\noindent Here, $\omega$ is constant and the subscript $0$ is
related to present time. Eq. (\ref{e28}) displays the relationship
between the energy density and the scale factor during reheating
as

\begin{equation}\label{e28}
\frac{\rho_{\mathrm{reh}}}{\rho_\mathrm{f}} =
(\frac{a_{\mathrm{reh}}}{a_\mathrm{f}})^{-3(1+\omega_{\mathrm{reh})}}.
\end{equation}

\noindent The relationship between the values of scale factor at
the end of inflation ($a_\mathrm{f}$), reheating ($a_{\mathrm{reh}}$), and the
reheating number of e-foldings ($N_{\mathrm{reh}}$) is

\begin{equation}\label{e29}
    N_{\mathrm{reh}} = \ln (\frac{a_{\mathrm{reh}}}{a_\mathrm{f}}),
\end{equation}

\noindent combined with Eq.~(\ref{e28}) to obtain

\begin{equation}\label{e30}
    N_{\mathrm{reh}} = \frac{1}{3(1+\omega_{\mathrm{reh}})}\ln(\frac{\rho_\mathrm{f}}{\rho_{\mathrm{reh}}}),
\end{equation}

\noindent which subsequently results in

\begin{equation}\label{e31}
    \rho_{\mathrm{reh}} = \rho_\mathrm{f} e^{-3(1+\omega_{\mathrm{reh}}){N_{\mathrm{reh}}}}.
\end{equation}

\noindent At the end of inflation, we get $\varepsilon_V=1$, which means that ${\dot{H}}=-{H^2}$. By inserting this equation in Eqs. (\ref{e7}) and (\ref{e8}), the link between $\rho_\mathrm{f}$ and $V(\phi_\mathrm{f})$ can be obtained as

\begin{equation}\label{e32}
    \rho_\mathrm{f} = \frac{3}{2} V(\phi_\mathrm{f}).
\end{equation}

\noindent Moreover, according to the Stefan–Boltzmann law, the relationship between $\rho_{\mathrm{reh}}$ and $T_{\mathrm{reh}}$ takes the form \cite{Kawai and Nakayama (2015), Haro and Saló (2023)}

\begin{equation}\label{e33}
    \rho_{\mathrm{reh}} = \frac{\pi^{2}}{30}g_{\mathrm{reh}} T_{\mathrm{reh}}^{4},
\end{equation}

\noindent in which $g_{\mathrm{reh}}$ denotes the number of relativistic
species at the end of reheating. $T_{\mathrm{reh}}$ is related to the
current temperature of the universe $T_{0}$ via

\begin{equation}\label{e34}
    T_{\mathrm{reh}} = T_{0}(\frac{a_{0}}{a_{\mathrm{reh}}})(\frac{43}{11 g_{\mathrm{reh}}})^{\frac{1}{3}}.
\end{equation}

\noindent Here,

\begin{equation}\label{e35}
    \frac{a_{0}}{a_{\mathrm{reh}}}=(\frac{a_{0}H_\mathrm{k}}{\mathrm{k}})e^{-N_e}e^{-N_{\mathrm{reh}}},
\end{equation}

\noindent where $\mathrm{k}={a_\mathrm{k}}{H_\mathrm{k}}$. Using Eqs.~(\ref{e31})-(\ref{e35}), $N_{\mathrm{reh}}$ is obtained as

\begin{eqnarray}\label{e36}
   && N_{\mathrm{reh}} = \frac{4}{3(1+\omega_{\mathrm{reh})}}\Big[\frac{1}{4} \ln(\frac{45}{\pi^{2}g_{\mathrm{reh}}})+\frac{1}{4}\ln
    ( \frac{V(\phi_\mathrm{f})}{H_\mathrm{k}^{4}})\nonumber\\ &&+\frac{1}{3} \ln(\frac{11 g_{\mathrm{reh}}}{43}) + \ln(\frac{\mathrm{k}}{a_{0}T_{0}}) + N_e + N_{\mathrm{reh}}\Big].
\end{eqnarray}

\noindent Now, by substituting ${g_{\mathrm{reh}}}=106.75$,
$\frac{\mathrm{k}}{a_0}=0.05$ Mpc$^{-1}$, and ${T_0}=2.725$ K into Eq.~(\ref{e36}) \cite{Dai et al. (2014), Mirtalebian et al. (2021), Cheong et al. (2022), Zhou et al. (2022)}, $N_{\mathrm{reh}}$ is
finally simplified as

\begin{eqnarray}\label{e37}
    N_{\mathrm{reh}} = \frac{4}{1-3\omega_{\mathrm{reh}}}\Big[61.6 - \frac{1}{4}\ln(\frac{V(\phi_\mathrm{f})}{H_{\mathrm{k}}^{4}})-{N_e}\Big].
\end{eqnarray}

\noindent Using $r=16 {\varepsilon_V}$, one reaches

\begin{equation}\label{e38}
    H_\mathrm{k} = M_{\mathrm{Pl}}\frac{\pi \sqrt{r {A_s}}}{\sqrt{2}},
\end{equation}

\noindent in which $A_s=2.196\times{10^{-9}}$ \cite{Dai et al. (2014), Mirtalebian et al. (2021), Cheong et al. (2022), Zhou et al. (2022)}. By employing Eqs.~(\ref{e31})-(\ref{e33}), $T_{\mathrm{reh}}$
emerges as

\begin{equation}\label{e39}
    T_{\mathrm{reh}} = \left(\frac{45}{\pi^{2}g_{\mathrm{reh}}}V(\phi_\mathrm{f})\right)^{0.25}e^{\frac{-3(1+\omega_{\mathrm{reh}}){N_{\mathrm{reh}}}}{4}}.
\end{equation}

\noindent Considering ${\varepsilon}_V\ll1$ during the primary inflationary era, $V(\phi_\mathrm{k})$ is calculated as

\begin{equation}\label{e40}
    V(\phi_\mathrm{k})=3 M_{\mathrm{Pl}}^{2} H_{\mathrm{k}}^{2},
\end{equation}

\noindent which leads to

\begin{eqnarray}\label{e41}
   && V({\phi_\mathrm{f}}) = V({\phi_\mathrm{k}}) \frac{-\cos (N {\phi_\mathrm{f}}) + \gamma \cos (2 N {\phi_\mathrm{f}})}{-\cos (N {\phi_\mathrm{k}})+\gamma \cos (2 N {\phi_\mathrm{k}})}
    =\nonumber\\ &&3M^{2}_{\mathrm{Pl}}H^{2}_{\mathrm{k}}\frac{-\cos (N {\phi_\mathrm{f}}) + \gamma \cos (2 N {\phi_\mathrm{f}})}{-\cos (N {\phi_\mathrm{k}})+\gamma \cos (2 N {\phi_\mathrm{k}})}.
\end{eqnarray}

\noindent Using Eqs.~(\ref{e37})-(\ref{e41}), it is a matter of calculations to reach reheating number of e-foldings (\ref{e23}), and reheating final temperature (\ref{e24}).\\

\section*{Acknowledgments}
The authors thank the anonymous referee for the valuable comments. The work of KB was supported by the JSPS KAKENHI Grant Numbers 21K03547, 24KF0100 and Competitive Research Funds for Fukushima University Faculty (25RK011).


\end{document}